\documentclass[powders,article,preprints,pdftex,moreauthors]{Definitions/mdpi_preprint} 
\firstpage{1} 
\pubvolume{1}
\issuenum{1}
\articlenumber{0}
\pubyear{2022}
\copyrightyear{2022}
\datereceived{} 
\dateaccepted{} 
\datepublished{} 
\hreflink{https://doi.org/} 
\pdfoutput=1

\usepackage{siunitx}
\usepackage{subcaption}
\usepackage{amssymb,amsfonts}
\usepackage{dsfont}
\usepackage{bbm}
\usepackage{verbatim}

\newcommand{\densityfeed}{f^\mathrm{f}}

\newcommand{\densityconcentrate}{f^\mathrm{c}}
\newcommand{\densitytailing}{f^\mathrm{t}}

\newcommand{\massParticlesConcentrate}{m^\mathrm{c}}
\newcommand{\massParticlesFeed}{m^\mathrm{f}}

\newcommand{\numberParticlesConcentrate}{n^\mathrm{c}}
\newcommand{\numberParticlesFeed}{n^\mathrm{f}}

\newcommand{\density}{\rho}

\newcommand{\R}{\mathbb{R}}
\newcommand{\Prob}{\mathbb{P}}
\newcommand{\Z}{\mathbb{Z}}

\newcommand{\aspectRatio}{\psi}
\newcommand{\AspectRatio}{\Psi}
\newcommand{\areaDiameter}{d_\mathrm{A}}
\newcommand{\AreaDiameter}{D_\mathrm{A}}
\newcommand{\minFeret}{d_\mathrm{min}}
\newcommand{\maxFeret}{d_\mathrm{max}}

\newcommand{\meanVolume}{\overline{V}}
\newcommand{\descriptor}[1]{x^{(#1)}}
\newcommand{\liste}[3]{#1{#2},\dots,#1{#3}}

\newcommand{\id}{\mathbbm{1}}

\newcommand{\CZero}{\mathrm{C_0}}
\newcommand{\CSix}{\mathrm{C_6}}
\newcommand{\CTen}{\mathrm{C_{10}}}
\DeclareMathOperator*{\argmin}{arg\,min}
\DeclareSIUnit{\molar}{M}

\Title{Application of multivariate Tromp functions for evaluating the joint impact of particle size, shape and wettability on the separation  of ultrafine particles via flotation}

\TitleCitation{Application of multivariate Tromp functions for evaluating the joint impact of particle size, shape and wettability on the separation  of ultrafine particles via flotation}

\Author{Johanna Sygusch $^{1,a,}$*\orcidA{}, Thomas Wilhelm$ ^{2,a,}$*\orcidB{}, Orkun Furat $^{2}$, Kai Bachmann $^{1}$, Volker Schmidt $^{2}$ and Martin Rudolph $^{1}$}
 
\AuthorNames{Johanna Sygusch, Thomas Wilhelm, Orkun Furat, Kai Bachmann, Volker Schmidt and Martin Rudolph}

\AuthorCitation{Sygusch, J.; Wilhelm, T.; Furat, O.; Bachmann, K.; Schmidt, V.; Rudolph, M.}

\address{%
$^{1}$ \quad Helmholtz-Zentrum Dresden-Rossendorf, Helmholtz Institute Freiberg for Resource Technology, \mbox{09599 Freiberg, Germany}
\\
$^{2}$ \quad Institute of Stochastics, Ulm University, 89069 Ulm, Germany 
\\
$^{a}$ \quad These authors contributed equally.
}

\corres{Correspondence:  j.sygusch@hzdr.de and thomas.wilhelm@uni-ulm.de}

\abstract{Froth flotation predominantly separates particles according to their differences in wettability. However, other particle properties such as size, shape or density significantly influence the separation outcome as well. Froth flotation is most efficient for particles within a size range of about $20-\SI{200}{\micro\meter}$, but challenges arise for very fine or coarse particles that are accompanied by low recoveries and poor selectivity. While the impact of particle size on the separation behavior in flotation is well-known by now, the effect of particle shape is less studied and varies based on the investigated zone (suspension or froth) and the separation apparatus used. Beyond these complexities, many particle properties are correlated, making it challenging to analyze the isolated impact of individual properties on the separation behavior. Therefore, a multidimensional perspective on the separation process, considering multiple particle properties, enhances the understanding of their collective influence. In this paper the two-dimensional case is studied, i.e.,  a parametric modeling approach is applied to determine bivariate Tromp functions from scanning electron microscopy-based image data of the feed and the separated fractions. With these functions it is possible to characterize the separation behavior of particle systems. Using a model system of ultrafine (<$\SI{10}{\micro\metre}$) particles, consisting of either glass spheres or glass fragments with different wettability states as the floatable and magnetite as the non-floatable fraction, allows for investigating the influence of descriptor vectors, consisting  of size, shape and wettability, on the separation. In this way, the present paper contributes to a better understanding of the complex interplay between certain property vectors for the case of ultrafine particles. Furthermore, it demonstrates the benefits of using multivariate Tromp functions for evaluating separation processes,  and points out the limitations of SEM based image measurements by means of mineral liberation analysis (MLA) for the studied particle size fraction.}


\begin{document}

\section{Introduction}

Many separation processes are designed focusing on a certain particle property that is the dominating feature for a successful separation, e.g., in flotation the particles are predominantly separated due to differences in their wettabilities. Particles that are hydrophobic attach to gas bubbles and are recovered through a froth, while  hydrophilic particles remain in suspension. However, in addition to this dominating separation feature, other particle properties also play an important role for the process outcome. In the case of flotation, apart from wettability, the particle size, shape, density or surface roughness significantly influence the separation. Regarding the particle size, there is a range of around $20-\SI{200}{\micro\meter}$ for which the separation by flotation works very efficiently~\cite{Trahar1981,Schubert1996,Wills2016}. However, if the particles are either too fine or too coarse the recovery as well as the selectivity decline significantly. The challenges of processing very fine particles are their unselective recovery by entrainment (for fine particles of the gangue material), which increases with decreasing particle size, and thus reducing the product grade. Additionally, there is a risk of slime coating onto coarser valuable particles, inhibiting their recovery ~\cite{Trahar1976,Miettinen2010,DeGontijo2007,Konopacka2010,Leistner2017}. Furthermore, very fine particles have rather slow flotation kinetics since the particle-bubble-collision efficiency depends strongly on the particle-bubble size ratio and thus decreases along with decreasing particle size~\cite{Dai2000}. Coarse particles, on the other hand, have a high probability of colliding with a bubble, but their particle-bubble aggregates are less stable and they can detach from the bubble more easily than finer particles, resulting in reduced recoveries~\cite{DeGontijo2007}.

Whereas many studies come to the same conclusion on how the size is affecting the separation by flotation, the influence of the particle shape is not as straightforward. This complexity arises from the zone under investigation, i.e., either suspension or froth zone. The choice of the separation apparatus, i.e., micro flotation, mechanical agitator-type froth flotation, column flotation etc., and the underlying flotation mechanism, whether true flotation or entrainment, further contribute to the variability in how the particle shape affects the separation. Investigations in which the froth zone was more or less not considered showed that the flotation of irregularly shaped particles and/or particles with rough surfaces is accompanied by higher recoveries and faster kinetics than if particles are used that are rather spherical and/or have a smooth surface. This is mainly supposed to be a result of the facilitated rupture of the liquid film between the bubble and an edgy/rough particle, resulting in shorter attachment times and higher attachment probabilities~\cite{Koh2009,Vaziri Hassas2016,Verrelli2014,Xia2017,Lu2005,Chen2022}. 
Regarding the froth zone, Kursun et al.~\cite{Kursun2006} reported higher recoveries for particles that were elongated and flat than for those that are spherical. On the other hand, Sygusch et al.~\cite{Sygusch2023} used a defined particle system with different shapes and tested those using a combination of mechanical and column flotation and compared these experiments to a benchmark mechanical cell. They showed that the influence of particle shape on the recovery and the selectivity varies depending on the apparatus used as well as on the wettability of the particles. Studies investigating the impact of particle shape on entrainment also report diverse results. Little et al.~\cite{Little2016} and Kupka et al.~\cite{Kupka2020} showed that for their particle systems the entrainment increased with increasing particle roundness. However, Wiese et al.~\cite{Wiese2015} and Sygusch et al.~\cite{Sygusch2023} reported that for their cases the entrainment was more pronounced for elongated particles and fragments, respectively. 

Although the wettability is the key separating feature for flotation, its effect on the process is usually only studied with respect to the suspension zone. Here, several studies report that the probability that a particle attaches to a bubble increases with its hydrophobicity. However, if the froth zone is considered, particles that are too hydrophobic induce bubble coalescence, resulting in the destabilisation of the froth, i.e., froth collapse, reducing the recovery, which is why for flotation usually particles with a moderate hydrophobicity are favored~\cite{Albijanic2010,Drelich2017,Johansson1992,Ata2003,Sygusch2023}.
Furthermore, many of these particle properties interact with each other. For example, the entrainment of particles is not only a function of their size but is also influenced by their mass density (as this affects their settling velocity). Not only the wettability influences the froth characteristics, but also the shape, as several studies showed that the critical contact angle varies for differently shaped particles~\cite{Aveyard1994, Dippenaar1982, Ata2012, Farrokhpay2011, Kaptay2012, Ata2004, Sygusch2023}.
Therefore, adopting a multidimensional view on the separation process, i.e., considering multiple particle properties  and descriptors rather than focusing on a single one, allows for a more comprehensive understanding of how these properties collectively influence the separation behavior of particles and at the same time reveals the interplay among these properties.

One way of obtaining this kind of multidimensional information is by analyzing the different material streams of the separation process, i.e., feed, concentrates and tailings, via automated mineralogy, from which particle discrete data is obtained, which is then used to determine multivariate Tromp functions. 
Previous studies have already proven this to be a valuable method for studying separation processes, as shown by Schach et al.~\cite{Schach2019} or Leißner et al.~\cite{Leissner2016}, who used kernel density estimates to characterize the separation according to particle size and mass density in a Falcon separator, as well as Wilhelm et al.~\cite{WilhelmSygusch2023}, who computed bivariate Tromp functions based on copulas to investigate the influence of particle size and shape on flotation. 
Other multivariate approaches, not based on Tromp functions, have also been presented in literature, for example Pereira et al.~\cite{Pereira2021a,Pereira2021b} developed a particle-tracking method that uses a regularized logistic regression model to obtain probability values for the behavior of individual particles, where several particle properties are considered.

In the present paper, we determine bivariate Tromp functions from scanning electron microscopy-based image data of the feed and the separated fractions by using a parametric modeling approach as in Wilhelm et al.~\cite{WilhelmSygusch2023}. This  involves the  fitting of (univariate) marginal densities of the individual particle descriptors, followed by the computation of an adequate copula density by utilizing Archimedean copulas~\cite{Nelsen2006,Furat2019}, which capture the dependencies between the particle descriptors. However, in contrast to~\cite{WilhelmSygusch2023}, image measurements are now available for all output streams (five concentrates and tailings), eliminating the need for employing the optimization approach considered in~\cite{WilhelmSygusch2023} to compute bivariate Tromp functions in the absence of measurements of the concentrate. The parametric modeling approach for computing bivariate Tromp functions is utilized to characterize the influence of particle descriptor vectors of shape and size as well as changes in particle wettability on the separation process. These findings are then connected with classical flotation results, including grade, recovery, as well as mass and water pull in flotation-based separation. Furthermore, an additional investigation on the entrainment behavior of ultrafine particles is proposed. By modifying the wettability of the valuable fraction (glass particles) information is obtained on the entrainment behavior of the hydrophilic magnetite (as the gangue) depending on the properties of the particles it is mixed with. Additionally, purely hydrophilic systems are tested (hydrophilic glass spheres or glass fragments as valuables and hydrophilic gangue) to investigate the influence of the particle shape on the entrainment of the valuable fraction as well.

The rest of this paper is structured as follows. Section~\ref{Sec:Materials and methods} deals with the materials and methods considered in this paper. In particular, in Section~\ref{Sec:Materials} a description of the particle systems used to prepare feed materials for the separation experiments is provided. Section~\ref{Sec:Flotation-based separation processes} outlines the flotation-based separation process. Then, Sections~\ref{Sec:Mineral liberation analysis} and 
\ref{Sec:Particle-based segmentation}
present details on the microscopy technique employed to generate image data for a quantitative analysis of the separation results. The stochastic modeling approach of descriptor vectors is stated in Section~\ref{Sec:Stochastic modeling of particle descriptor vectors and computation of multivariate Tromp functions} and the computation of bivariate Tromp functions is explained   in Sections~\ref{Sec:Computation of yield} and \ref{Sec:Probability densities of descriptor vectors associated with particles in the feed and concentrate in flotation separation}. Conditional univariate Tromp functions, conditioned on particle size and shape classes, are considered in Section~\ref{Sec:Univariate Tromp functions conditioned on particle size and shape classes}. This is followed by 
Section~\ref{Sec:Results and discussion}, which comprises the results and their discussion, including classical flotation results in Section~\ref{Sec:Classic flotation results} as well as results related to the influence of particle size and shape on the entrainment of ultrafine particle in Section~\ref{Sec:Influence of particle size and shape on the entrainment of ultrafine particles}. Additionally, it explores the combined influence of particle size, shape and wettability of ultrafine particles in Section~\ref{Sec:Influence of particle shape, size and wettability on the separation behavior of ultrafine particles}. Furthermore, in Section~\ref{Sec:General discussion} a general discussion on the usability of results obtained from image measurements is given. Finally, Section~\ref{Sec:Conclusion} concludes.


\section{Materials and methods}\label{Sec:Materials and methods}


\subsection{Materials}\label{Sec:Materials}

\begin{figure}[ht!]
\captionsetup[subfigure]{labelformat=empty}
        \centering
    \begin{subfigure}[b]{0.75\textwidth}
      \includegraphics[scale=0.75]{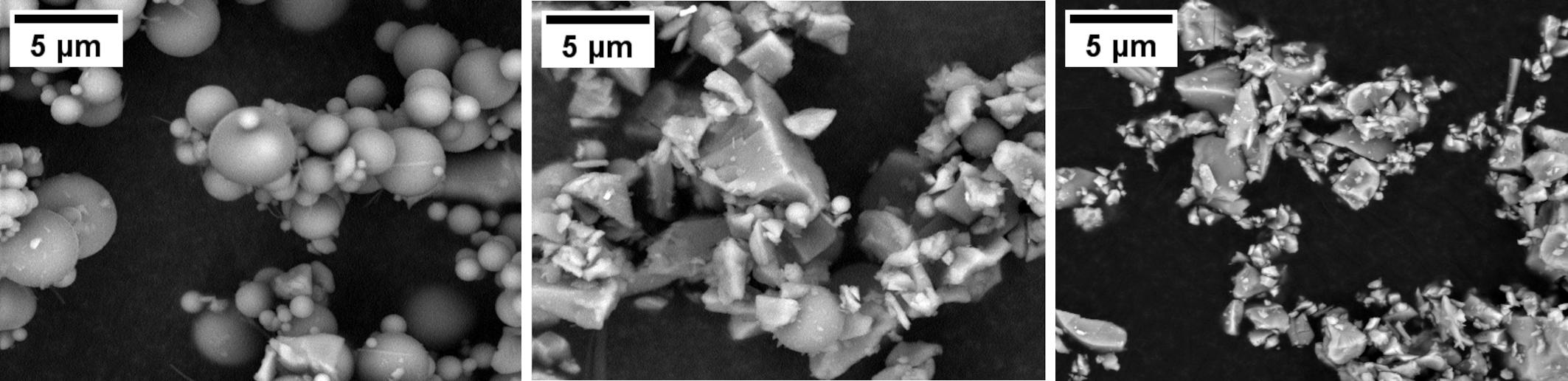}
           \end{subfigure}
\caption{SEM images of glass spheres (left), glass fragments (middle) and magnetite (right).}
    \label{Fig:SEM_images}
\end{figure}

\begin{figure}[ht!]
\captionsetup[subfigure]{labelformat=empty}
        \centering
         \begin{subfigure}[b]{0.77\textwidth}
    \includegraphics[scale=1.5]{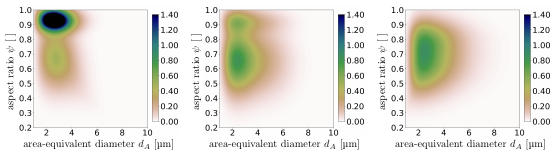}
     \end{subfigure}
    \caption{Bivariate probability densities representing  particle descriptors of shape (aspect ratio) and size (area-equivalent diameter) for glass spheres (left), glass fragments (middle), and magnetite (right). The computation of these bivariate probability densities is based on the copula-based approach outlined in~\cite{WilhelmSygusch2023}. The particle descriptors are obtained from image analysis by means of  mineral liberation 
analysis (MLA), in which the particle fractions are analyzed individually.}
    \label{Fig:Bivariate densities}
\end{figure}

Figure~\ref{Fig:SEM_images} shows scanning electron microscopy  (SEM) images of the particles that are used as feed material for the flotation experiments, where glass particles with different shapes, spheres and fragments, are used as the floatable fraction and magnetite is used as the non-floatable fraction. Ultrafine size fractions of magnetite have been purchased from Kremer Pigmente, Germany, and analysis via X-ray diffraction confirmed its purity. Glass spheres and fragments both consist of soda-lime glass and have been purchased from VELOX, Germany, as SG7010 and SG3000, respectively. The glass spheres considered in this study have particle sizes below $\SI{10}{\micro\metre}$ (SG7010). Ultrafine glass fragments are obtained by milling and aero classification of coarser glass spheres (SG3000). The bivariate probability densities of pairs of particle descriptors, namely the area-equivalent diameter and the aspect ratio given by means of Equations~(\ref{Eq:Definition areaDiameter}) and~(\ref{Eq:Definition aspectRatio}), are visualized in Figure~\ref{Fig:Bivariate densities}. 

Both glass particle fractions have a mass density of $\SI[per-mode=symbol]{2500}{\kilogram\per\meter^3}$ with a stationary settling velocity of v-glass equal to $8.27 \times 10^{-6}\SI[per-mode=symbol]{}{\meter\per\second}$. Magnetite has a mass density of $\SI[per-mode=symbol]{5200}{\kilogram\per\meter^3}$ and, with a v-magnetite of $2.31 \times 10^{-5}\SI[per-mode=symbol]{}{\meter\per\second}$, a faster settling velocity (calculation of the settling velocity is based on spherical particles for the Stokes regime~\cite{Schubert1989}). While the magnetite was used as received, the glass particles underwent an esterification reaction using n-alcohols, which allows for generating particle fractions with defined wettability states depending on the alkyl chain length of the alcohol used, as  presented in~\cite{SyguschRudolph2021}. Three different wettability states of glass particles are used for this study: (i) pristine, unesterified hydrophilic particles,  (ii) particles that are hydrophobized using the primary alcohols 1-hexanol ($\CSix$, Carl Roth $\geq$ $98\%$, used as received), and (iii) 1-decanol ($\CTen$, Carl Roth $\geq$ $99\%$, used as received) resulting in esterified particles with moderate and strong hydrophobicities. Table~\ref{Table:ContactAngles} displays the respective contact angles that increase with increasing hydrophobicity, measured on equally treated glass slides.

\begin{table}[ht!]
\begin{center}
\caption{Static contact angles of glass slides in their pristine unesterified state ($\CZero$), esterified with 1-hexanol ($\CSix$) and 1-decanol ($\CTen$), measured via the sessile drop method using water. The glass slides have the same chemical composition as the glass particles and were treated identically.}
\begin{tabular}{||c c||} 
 \hline
 wettability experiment & contact angle in $^{\circ}$ \\ [0.5ex] 
 \hline\hline
 $\CZero$ & $38.3 \pm 0.6$ \\ 
 \hline
 $\CSix$ & $87.3 \pm 1.2$ \\
 \hline
 $\CTen$ & $105.3 \pm 0.4$ \\
 \hline
\end{tabular}
\label{Table:ContactAngles}
\end{center}
\end{table}


\subsection{Flotation-based separation experiments}\label{Sec:Flotation-based separation processes}

\begin{figure}[ht!]
\centering
      \includegraphics[scale=0.5]{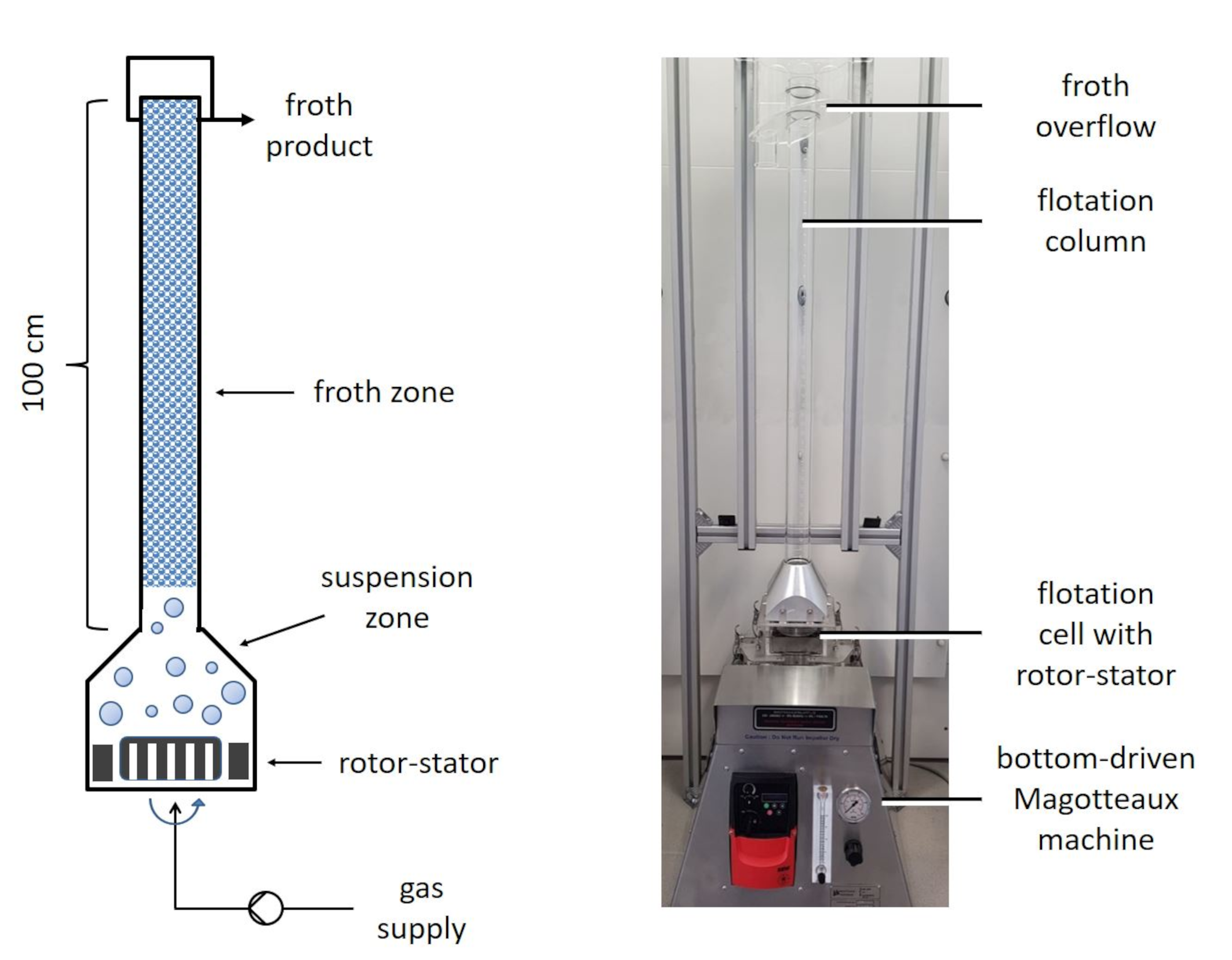}
\caption{Schematic diagram (left) and the actual lab set-up (right) of the \textit{MultiDimFlot} separation apparatus used for the flotation experiments in this study.}
\label{Fig:Separation_apparatus}
\end{figure}

All flotation experiments have been carried out using the newly developed \textit{MultiDimFlot} separation apparatus, shown in Figure~\ref{Fig:Separation_apparatus}, which combines mechanical agitator-type froth flotation using a bottom-driven Magotteaux machine ($\SI{12}{\centi\metre}$ x $\SI{12}{\centi\metre}$) with column flotation, where  a column length of $\SI{100}{\centi\metre}$ and a $\SI{5}{\centi\metre}$ diameter was used.

The flotation experiments were conducted at a rotational speed of $\SI{600}{\per\minute}$ using an airflow rate of $\SI[per-mode=symbol]{0.9}{\l\per\min}$ and a superficial gas velocity of $\SI[per-mode=symbol]{0.76}{\cm\per\second}$. All experiments are conducted in batch mode using $4.8 \%$ (w/w) pulp density with the glass particles and magnetite in a weight ratio of $1:9$, respectively. Poly(ethylenglycol) (PEG, Carl Roth with a molecular weight of $\SI[per-mode=symbol]{10000}{\g\per\mol}$) is used as frother. No collector is used for flotation, since the glass particle wettability is modified prior to flotation, hence no conditioning is required.
The particles are dispersed in a $10^{-2}\, \SI{ }{\molar}$ KCl background solution with a PEG concentration of $10^{-5}\, \SI{ }{\molar}$ using an Ultra Turrax (dispersion tool S25N-25F) from IKA, Germany, for $\SI{1}{\minute}$ at $\SI{11000}{\per\minute}$ , resulting in a dispersion with a pH of $9$. Flotation experiments are carried out for $\SI{8}{\minute}$ with concentrates being taken after $ 1, 2, 4, 6$, and $\SI{8}{\minute}$ by scrapping off the froth every \SI{10}{\second}. The concentrates and tailings are dewatered via centrifugation and dried in a drying cabinet. Characterization includes gravimetric analysis for mass balancing, X-ray fluorescence (with the Bruker S1 TITAN handheld device) to obtain the chemical composition, laser diffraction (HELOS,  Sympatec) for the particle size and mineral liberation analysis to obtain the composition of particle systems and particle-discrete information on size and shape.


\subsection{Sample preparation and SEM-based automated mineralogy}\label{Sec:Mineral liberation analysis}

Polished blocks were prepared for analysis by mixing representative sample splits with graphite powder, embedding the mixture in epoxy resin, followed by slicing, rotating and remounting to reduce the effects of gravity settling~\cite{Heinig2015}. The MLA was conducted at the Helmholtz Institute Freiberg for Resource Technology, utilizing a ThermoFisher (formerly FEI) Quanta 650F MLA system equipped with two Bruker Quantax X-Flash 5030 energy-dispersive X-ray spectroscopy (EDS) detectors. To generate false color images, backscattered electron (BSE)   images and EDX analyses were seamlessly integrated using FEI's MLA suite, version 3.1.4. In the BSE images, the gray scale's lower limit was established at epoxy resin ($<20$), while the upper limit was set to copper ($245-255$). A comprehensive characterization of each mineral phase and its distribution across the entire samples was accomplished by mapping with one EDX measurement point per defined grain. This mapping was conducted using \textit{Extended BSE liberation analysis} (XBSE mode~\cite{Fandrich2007,Schulz2020}), providing high resolution and enabling analysis post-BSE image segmentation based on user-defined parameters, such as minimum grain size. The specific measurement parameters were configured as follows: an acceleration voltage of $\SI{15}{\kilo\volt}$, a probe current of $10$ nA, a horizontal field width of $\SI{250}{\micro\meter}$, and a frame resolution of $1000$ pixels. These settings resulted in a spatial resolution of $0.25$ microns per pixel. Further details regarding the measurement procedure can be found in Bachmann et al.~\cite{Bachmann2017}.


\subsection{Particle-based segmentation}\label{Sec:Particle-based segmentation}

Using 2D image data obtained by MLA, certain image processing steps are applied to each image measurement. First, all particles from a specific material (glass or magnetite) are extracted by a phase-based segmentation. In MLA measurements, this segmentation is simply obtained by extracting all regions that have the same label, i.e., corresponding to a specific material. Furthermore, challenges related to particle agglomeration are encountered, making particle-wise segmentation challenging. To address this, the watershed algorithm~\cite{Roerdink2001, Soille2003} was employed on all segmented regions observed in the image measurements corresponding to agglomerates, where one particle is connected with one or more other particles. A common issue with the watershed algorithm is oversegmentation, where single particles are often divided into multiple segments. However, in the context of the present paper, oversegmentation is avoided by assuming that all particles in the systems of spheres, fragments, and magnetite are convex. Consequently, the watershed algorithm is applied only to those regions observed in the image measurements where the ratio of area (number of pixels belonging to the corresponding region) to the area of the  convex hull (number of pixels belonging to the smallest convex area containing the corresponding region) is smaller than $0.7$. For single particles that are not part of an agglomerate, this ratio is larger than $0.7$ due to  particle convexity, whereas for agglomerates, this ratio is typically smaller than $0.7$. This ratio is also known as convexity factor~\cite{Chiu2013} and is applied to similar particle systems in~\cite{Furat2018}.

\begin{table}[ht!]
\caption{The number of particles observed in the MLA images for various particle systems, different wettability experiments ($\CZero$, $\CSix$, $\CTen$), and for concentrate or tailings. The corresponding number of particles excluded after particle-based segmentation (where the area-equivalent diameter of the particles is smaller than $\SI{1}{\micro\meter}$ or larger than $\SI{10}{\micro\meter}$) is provided in brackets.
 \label{Tab:Number of observed and excluded particles}}
		\newcolumntype{B}{>{\centering}p{0.2\textwidth}}
  		\newcolumntype{D}{>{\raggedright}p{0.06\textwidth}}
      	\newcolumntype{E}{>{\centering}p{0.02\textwidth}}
		\newcolumntype{A}{>{\raggedright}p{0.2\textwidth}}
		\begin{tabularx}{\textwidth}{B|DCBC}
			\toprule
			& &&wettability experiment& \\
    particle system & &$\CZero$&$\CSix$&$\CTen$ \\
    \midrule
    spheres & concentrate: &$252522 \,(3128)$
                            &$402105 \,(3877)$
                            &$366878 \,(3015)$ \\
             & tailing:     &$18223 \,(163)$
                            &$5715 \,(80)$
                            &$3114 \,(96)$ \\
    \midrule
    fragments & concentrate: &$299724 \,(2131)$
                            &$463711 \,(3017)$
                            &$139505 \,(847)$\\
             & tailing:     &$12677 \,(220)$
                            &$15593 \,(158)$
                            &$22623 \,(200)$\\
             \midrule
magnetite \textbackslash spheres 
            & concentrate:  &$263301 \,(11831)$
                            &$137235 \,(6756)$
                            &$20252 \,(1102)$\\
             & tailing:     &$220400 \,(3302)$
                            &$118529 \,(4292)$
                            &$130600 \,(4443)$\\
                            \midrule
magnetite \textbackslash fragments             
            & concentrate:  &$565991 (8426)$
                            &$326359 \,(7847)$
                            &$28028 \,(896)$\\
             & tailing:     &$121782 (4646)$
                            &$196805 \,(2892)$
                            &$236174 \,(3575)$ \\
			\bottomrule
		\end{tabularx}
\end{table}

After the extraction of similarly labeled regions and the application of the watershed algorithm to account for agglomerates, a particle-based segmentation is obtained. To further refine the segmentation and exclude undesirable artifacts, all extracted regions with an area-equivalent diameter smaller than $\SI{1}{\micro\meter}$ or larger than $\SI{10}{\micro\meter}$ are disregarded by means of Equation~(\ref{Eq:Definition areaDiameter}). The obtained results are presented in Table~\ref{Tab:Number of observed and excluded particles}, providing information on the number of particles observed in the MLA images and the subsequent exclusion after particle-based segmentation. This data encompasses various particle systems and includes results from different wettability experiments involving both concentrate and tailings.

\subsection{Stochastic modeling of particle descriptor vectors for the computation of multivariate Tromp functions}\label{Sec:Stochastic modeling of particle descriptor vectors and computation of multivariate Tromp functions}

A system of particles, observed by image measurements and extracted by   particle-based segmentation, can be described by a set of descriptor vectors. In this context, MLA images provide a planar section of a three-dimensional particle system within a certain sampling window $W\subset\Z^2$. In the present paper, based on the particle-wise segmentation of  2D images, each particle is represented by a $2$-dimensional descriptor vector $x = (x_1, x_2)\in\R^2$, where the first entry $x_1$ of $x$ denotes the particle's size, while the second entry $x_2$ characterizes the particle's shape. In order to evaluate the separation behavior of particle systems in separation experiments, the entirety of particle descriptor vectors associated with particles of the feed material and of the concentrate are modeled by number-weighted bivariate probability densities $\densityfeed:\R^2\rightarrow[0,\infty)$ and $\densityconcentrate:\R^2\rightarrow[0,\infty)$, respectively. This allows computing the number-weighted bivariate Tromp function $T:\R^2\to[0,1]$~\cite{Schach2019, WilhelmSygusch2023} given by
\begin{linenomath}
\begin{equation}\label{Eq:Number weighted Tromp function}
    T(x) = \begin{cases}\displaystyle
    \frac{\numberParticlesConcentrate}{\numberParticlesFeed}\,\frac{\densityconcentrate(x)}{\densityfeed(x)}, & \text{if } \densityfeed(x)>0,\\
    0, & \text{ if } \densityfeed(x)=0,
    \end{cases}
\end{equation}
\end{linenomath}
for each $x\in \R^2$, where $\numberParticlesConcentrate$ and $\numberParticlesFeed$ denote the number of particles in the concentrate and the feed, respectively. The value $T(x)$ of the Tromp function can be interpreted as the probability of a particle with descriptor vector $x$, to be separated into the concentrate. 

In the following a more detailed description of the considered size and shape particle descriptors is provided, see Section~\ref{Sec:Characterization of particles with size and shape descriptors}. Then, in Sections~\ref{Sec:Univariate stochastic modeling of single particle descriptors} and~\ref{Sec:Bivariate stochastic modeling of pairs of particle descriptors using Archimedean copulas}, methods for modeling the distribution of single particle descriptors by univariate probability densities and the distribution of pairs of descriptors by means of a parametric copula-based procedure are discussed.


\subsubsection{Characterization of particles by means of size and shape descriptors}\label{Sec:Characterization of particles with size and shape descriptors}
For particles observed in a planar section of a three-dimensional particle system, it is possible to compute various size and shape descriptors using the particle-wise segmentation of 2D images obtained by MLA measurements within some sampling window $W\subset\Z^2$, see Section~\ref{Sec:Mineral liberation analysis}. The size and shape descriptors used in the present paper are adopted from~\cite{WilhelmSygusch2023}.

In order to characterize the size of a particle's cross-section $P$ observed within the sampling window $W$, the area-equivalent diameter $\areaDiameter(P)$ of $P$ is determined, which is given by
\begin{linenomath}
\begin{equation}\label{Eq:Definition areaDiameter}
	\areaDiameter(P)=2 \sqrt{\frac{A(P)}{\pi}},
\end{equation}
\end{linenomath}
where $A(P)$ denotes the area of $P$. Note that $A(P)$ is computed from image data by counting the number of pixels belonging to the correspondingly discretized particle cross-section $P\subset W$. Recall that in the present study, particles with area-equivalent-diameter smaller than $\SI{1}{\micro\meter}$ are excluded from the analysis of image measurements due to limitations in the resolution of MLA.

Furthermore, the so-called minimum and maximum Feret diameters $\minFeret(P)$ and $\maxFeret(P)$ of $P$ are determined, by deploying the algorithm given in \cite{Hilsenstein2022}. More precisely, $\minFeret(P)$ and $\maxFeret(P)$ are the smallest and largest edge lengths of a minimum rectangular bounding box $B(x^\ast,y^\ast,\alpha^\ast,\beta^\ast,\theta^\ast)$ of $P$. Such a bounding box can be determined by solving the minimization problem
\begin{linenomath}
\begin{equation}\label{Eq:Definition Feret diameter}
    (x^\ast,y^\ast,\alpha^\ast,\beta^\ast,\theta^\ast) = \argmin_{\substack{
	(x,y,\alpha,\beta,\theta)\in \R^4 \times [0, \pi),\\ 0<\alpha\leq \beta,\\ P\subset B(x,y,\alpha,\beta,\theta) } 
	}
\alpha \cdot \beta,
\end{equation}
\end{linenomath}
where $B(x,y,\alpha,\beta,\theta)$ denotes a rectangle with edge lengths  $\alpha,\beta>0$ such that $\alpha\leq \beta$, which is rotated by $\theta \in [0,\pi)$ around its center $(x,y)\in \R^2$. Then, the minimum and maximum Feret diameters of $P$ are given by $\minFeret(P)=\alpha^\ast$ and $\maxFeret(P) =\beta^\ast$, respectively. This provides the aspect ratio $\aspectRatio(P)$ of $P$, which is given by
\begin{linenomath}
\begin{equation}\label{Eq:Definition aspectRatio}
	\aspectRatio(P)=\frac{\minFeret(P)}{\maxFeret(P)}\,.
\end{equation}
\end{linenomath}
Note that the aspect ratio $\aspectRatio$ defined in Equation~(\ref{Eq:Definition aspectRatio}) is a shape descriptor which allows to distinguish between elongated ($\aspectRatio(P)\ll 1$) and non-elongated particles ($\aspectRatio(P) \approx 1$). Analogously to the computation of the area of a particle cross-section from image data, the minimum and maximum Feret diameters $\minFeret(P)$ and $\maxFeret(P)$ of $P$ are determined by rescaling their values with the pixel size.


\subsubsection{Univariate stochastic modeling of single particle descriptors}\label{Sec:Univariate stochastic modeling of single particle descriptors}

The particle system observed in image measurements after a particle-wise segmentation can be characterized by a sample of particle descriptor vectors, where each descriptor vector is assigned to a single particle cross-section $P_i\subset W$ for $i=1,2,\dots,N$. Here, $N>0$ denotes the number of particle cross-sections in $W$. Each descriptor vector contains the particle size and shape descriptors as introduced in Section~\ref{Sec:Characterization of particles with size and shape descriptors}. Thus, in this study, the focus lies on determining two-dimensional descriptor vectors $\liste{\descriptor}{1}{N} \in \R^2$. As already mentioned above, the first entry of these descriptor vectors is the area-equivalent diameter of the corresponding particle cross-section, while the second entry describes the particle's shape using the aspect ratio of its planar cross-section. Thus, formally, the descriptor vectors of a  particle cross-sections are given by $x^{(i)} = (\areaDiameter(P_i),\aspectRatio(P_i))$ for $i=1,\dots,N$.

A univariate probability density is fitted from a parametric family $\{f_\theta \colon \theta \in \Theta \}$ of probability densities $f_\theta \colon \R \to [0,\infty)$ to each entry of the particle descriptor vectors (e.g.,  densities of normal, log-normal, gamma, or beta distributions), where $\Theta$ is the set of admissible parameters, see Table~\ref{Tab:Probability densities families}. The best fitting density and the corresponding parameters are chosen by means of the maximum-likelihood method~\cite{Held2014}.

\begin{table}[ht!]
\caption{Parametric families of univariate probability densities $f_\theta \colon \R \to [0,\infty)$ used later on in Sections~\ref{Sec:Influence of particle size and shape on the entrainment of ultrafine particles} and~\ref{Sec:Influence of particle shape, size and wettability on the separation behavior of ultrafine particles} for the fitting of bivariate densities. 
\label{Tab:Probability densities families}}
		\newcolumntype{B}{>{\centering}p{0.2\textwidth}}
		\newcolumntype{A}{>{\raggedright}p{0.45\textwidth}}
		\begin{tabularx}{\textwidth}{BAC}
			\toprule
			parametric family & probability density& \\
			\midrule
            normal & $f_{\theta}(x) = \frac{1}{\sqrt{2\pi\sigma^2}}\mathrm{e}^{-\frac{(x-\mu)^2}{2\sigma^2}}, $&$\theta=(\mu,\sigma)\in\R\times(0,\infty)$  \\ 
                log-normal & $f_{\theta}(x) = \frac{1}{\sqrt{2\pi\sigma^2x^2}}\mathrm{e}^{-\frac{(\log(x)-\log(\mu))^2}{2\sigma^2}}\id_{(0,\infty)}(x), $&$\theta=(\mu,\sigma)\in\R\times(0,\infty)$  \\ 
                gamma &$f_{\theta}(x) =\frac{1}{b^k\Gamma(k)}x^{k-1}\mathrm{e}^{\big(-\frac{x}{b}\big)}\id_{(0,\infty)}(x), $&$\theta=(k,b)\in(0,\infty)^2$  \\
                beta &$ f_{\theta}(x) = \frac{\Gamma(\alpha+\beta)}{\Gamma(\alpha)\Gamma(\beta)}x^{\alpha-1}(1-x)^{\beta-1}\id_{(0,1)}(x),  $&$\theta=(\alpha,\beta)\in(0,\infty)^2$  \\
			\bottomrule
		\end{tabularx}
\end{table}

Analogously to~\cite{WilhelmSygusch2023}, the characterization of the particle systems presented in this paper involves the consideration of bimodal probability densities as a convex combination $f_{\theta_1,\theta_2,w}=wf_{\theta_1}+(1-w)f_{\theta_2}$ of unimodal probability densities $f_{\theta_1},f_{\theta_2}:\R\to(0,\infty)$ for some $\theta_1,\theta_2\in\Theta$, where $w\in(0,1)$ is a mixing parameter. Introducing such bimodal densities allows for improved fits, albeit at the expense of increasing the number of model parameters. The best fitting distribution is selected according to the Akaike information criterion~\cite{Akaike1998}.


\subsubsection{Bivariate stochastic modeling of pairs of particle descriptors using Archimedean copulas}\label{Sec:Bivariate stochastic modeling of pairs of particle descriptors using Archimedean copulas}

The approach for modeling the distribution of individual particle descriptors, as explained in Section~\ref{Sec:Univariate stochastic modeling of single particle descriptors}, does not account for the correlation between these descriptors. To obtain a more comprehensive probabilistic representation of the observed particle system, bivariate probability densities are fitted to the dataset of pairs of two-dimensional descriptor vectors, which are computed as described in Section~\ref{Sec:Characterization of particles with size and shape descriptors}. 

In this study, the focus lies specifically on parametric families of Archimedean copulas, analogously to the approach considered in~\cite{WilhelmSygusch2023}. These families include the Clayton, Frank, Gumbel, and Joe copulas, as well as rotated versions of these copula families~\cite{Nelsen2006, Joe2014}, see Table~\ref{Tab:Copula.densities}. The bivariate probability density $f:\R^2\to[0,\infty)$ of a two-dimensional particle descriptor vector can be written in the form
\begin{linenomath}
\begin{equation}\label{Eq:Bivariate_pdf}
    f(x) = f_1(x_1)f_2(x_2)c(F_1(x_1),F_2(x_2))\qquad \mbox{for each  $x=(x_1,x_2)\in\R^2$, }
\end{equation}
\end{linenomath}
where $f_1,f_2:\R\to[0,\infty)$ denote the (univariate) marginal densities corresponding to $f$ and $c:[0,1]^2\to[0,\infty)$ is a bivariate copula density, i.e., a bivariate probability density with uniform marginal distributions on the unit interval $[0,1]$, see e.g.~\cite{Nelsen2006}. The best fitting bivariate density $f$ for a sample of  particle descriptor vectors obtained from image measurements is determined using maximum-likelihood estimation and the Akaike information criterion. This approach has been applied for parametric stochastic modeling of similar types of particle-discrete image data in~\cite{Furat2019, Ditscherlein2020}.

\begin{table}[ht!]
\caption{Parametric copula families used later on in Sections~\ref{Sec:Influence of particle size and shape on the entrainment of ultrafine particles} and~\ref{Sec:Influence of particle shape, size and wettability on the separation behavior of ultrafine particles}
for fitting the bivariate probability densities. 
\label{Tab:Copula.densities}}
		\newcolumntype{B}{>{\centering}p{0.2\textwidth}}
		\newcolumntype{A}{>{\raggedright}p{0.55\textwidth}}
		\begin{tabularx}{\textwidth}{BAC}
			\toprule
			parametric family & copula density &\\
			\midrule
                Clayton & $ c_{\theta}(u_1,u_2) = (\theta+1)(u_1u_2)^{-(\theta-1)}(u_1^{-\theta}+u_2^{-\theta}-1)^{-\frac{2\theta+1}{\theta}}, $&$\theta\in(0,\infty)$\\
               Frank &$     c_{\theta}(u_1,u_2) = \frac{\theta(1-\mathrm{e}^{-\theta})\mathrm{e}^{-\theta(u_1+u_2)}}{1-\mathrm{e}^{-\theta}-(1-\mathrm{e}^{-\theta u_1})(1-\mathrm{e}^{-\theta u_2})}, $&$\theta\in(0,\infty)$\\
               Gumbel &$     c_{\theta}(u_1,u_2) = \frac{\partial^2}{\partial u_1\partial u_2}e^{-((-\log(u_1))^\theta+(-\log(u_2))^\theta)^{\frac{1}{\theta}}}, $&$\theta\in(1,\infty)$\\
                Joe &$     c_{\theta}(u_1,u_2) = 1-(1-(1-(1-u_1)^{\theta}(1-(1-u_2)^{\theta})))^{1-\theta}, $&$\theta\in[1,\infty)$\\
			\bottomrule
		\end{tabularx}
\end{table}


\subsection{Computation of yield}\label{Sec:Computation of yield}

In order to compute bivariate Tromp functions as given in Equation~(\ref{Eq:Number weighted Tromp function}), it is essential to determine the yield, which is the ratio $\frac{\numberParticlesConcentrate}{\numberParticlesFeed}$ of the number of particles in the concentrate ($\numberParticlesConcentrate$) to the number of particles in the feed ($\numberParticlesFeed$). Note that this ratio cannot be directly obtained from image measurements, since these measurements only provide a statistically representative sample of the particle system. Therefore, it is utilized that the total mass $\massParticlesFeed$ of particles in the feed, which can be approximated by the number of particles $n^\mathrm{f}$ times the expected mass of a particle in the feed, which is given by $\int_{\R^2} m(x) \densityfeed(x) \,\mathrm{d}x$, i.e., $\massParticlesFeed \approx n^\mathrm{f}\int_{\R^2} m(x) \densityfeed(x) \,\mathrm{d}x$. Here, $m:\R^2\rightarrow [0,\infty)$ is a function which maps descriptor vector $x\in\R^2$ of particle cross-sections onto the particle mass $m(x)$, see e.g.,~\cite{Frank2019}. Analogously, for the concentrate one obtains $\massParticlesConcentrate \approx n^\mathrm{c}\int_{\R^2} m(x) \densityconcentrate(x) \,\mathrm{d}x$. Using these approximations of the total masses of particles, it is obtained that 
\begin{linenomath}
\begin{equation}\label{Eq:Yield}
\frac{\numberParticlesConcentrate}{\numberParticlesFeed} \approx \frac{{\massParticlesConcentrate}}{\massParticlesFeed}\frac{\int_{\R^2} m(x) \densityfeed(x)\,\mathrm{d}x}{\int_{\R^2} m(x) \densityconcentrate(x)\,\mathrm{d}x} = \frac{{\massParticlesConcentrate}}{\massParticlesFeed}\frac{\meanVolume^\mathrm{f}\density}{\meanVolume^\mathrm{c}\density},
\end{equation}
\end{linenomath}
where the expected masses of particles in feed and concentrate can be computed by the product of the mass density $\density$ of particles and their expected volume, denoted by $\meanVolume^\mathrm{f}$ for feed and $\meanVolume^\mathrm{c}$ for concentrate. Note that the total mass ratio $\massParticlesConcentrate / \massParticlesFeed$ and the mass density $\rho$ of particles are provided by the experimental setup. In addition, the expected volume of particles in feed and concentrate is obtained from image measurements. The volume of a single particle is considered to be equal to the volume of a sphere experiencing the same area-equivalent diameter as the particle under consideration. The expected volume of the particle system in question is determined by computing the mean volume across all particles under consideration.


\subsection{Probability densities of descriptor vectors associated with particles in the feed and concentrate}\label{Sec:Probability densities of descriptor vectors associated with particles in the feed and concentrate in flotation separation}

In flotation separation processes, the concentrate is often composed of multiple concentrate streams as described in Section~\ref{Sec:Flotation-based separation processes}. To obtain information on the entire concentrate, it is necessary to combine the information from all these individual concentrate streams. This means that the probability density $\densityconcentrate$ of descriptor vectors associated to particles in the concentrate is expressed as a convex combination of the probability densities of particle descriptor vectors obtained from image measurements of each individual concentrate stream.

In this case, the concentrate consists of five different concentrate streams. For each of these concentrate streams, the probability density $\densityconcentrate_i:\R^2\rightarrow[0,\infty)$ of descriptor vectors associated to the particles in the respective concentrate stream is computed, where $i=1,2,\dots, 5$. This computation is done using the methods described in Sections~\ref{Sec:Univariate stochastic modeling of single particle descriptors} and~\ref{Sec:Bivariate stochastic modeling of pairs of particle descriptors using Archimedean copulas}. The probability density $\densityconcentrate:\R^2\to[0,\infty)$ is then given by 
\begin{linenomath}
    \begin{align}
    \densityconcentrate (x) = \lambda_1 \densityconcentrate_1(x) + \dots + \lambda_5 \densityconcentrate_5(x)\qquad\mbox{for each $x\in\R^2$.}
\end{align}
\end{linenomath}
 The coefficients $\lambda_1,\ldots,\lambda_5\in[0,1]$ are given by
\begin{align*}
    \lambda_i = \frac{\numberParticlesConcentrate_i}{\numberParticlesConcentrate}\qquad\mbox{for $i\in\{1,\ldots,5\}$},
\end{align*}
where $n_i^\mathrm{c}$ is the number of particles in the $i$-th concentrate and $n^\mathrm{c}=\sum_{i=1}^5 n_i^\mathrm{c}$. Note that $n_i^\mathrm{c}$ is approximated by dividing the total mass of particles in the $i$-th concentrate by the expected mass of particles, analogously to the computation of the yield in Section~\ref{Sec:Computation of yield}. 

The computation of the probability density $\densityfeed$ of descriptor vectors associated to particles in the feed is performed as outlined in~\cite{WilhelmSygusch2023}. More precisely, the particle systems of the entire concentrate and the tailings need to be characterized in order to compute $\densityfeed$ as a convex combination of the probability densities $\densityconcentrate$ and $\densitytailing$, where $\densitytailing$
denotes the density of descriptor vectors associated with the particles in the tailings
~\cite{Buchmann2018}. In other words, $\densityfeed$ is given by
\begin{linenomath}
\begin{equation}\label{Eq:Feed probability density as convex combination}
    \densityfeed(x) = \frac{\numberParticlesConcentrate}{\numberParticlesFeed} \densityconcentrate(x) + (1-\frac{\numberParticlesConcentrate}{\numberParticlesFeed}) \densitytailing(x)\qquad\mbox{for each  $x\in\R^2$.} 
\end{equation}
\end{linenomath}

This approach is used to avoid numerical instabilities that can arise when computing Tromp functions by means of Equation~(\ref{Eq:Number weighted Tromp function}) and using $\densityfeed$ obtained from image measurements of the feed. These instabilities are due to the sensitivity of the Tromp function when computing denominator values in  Equation~(\ref{Eq:Number weighted Tromp function}) being close to zero. Note that this problem occurs when there are relatively few particles with certain descriptor vectors within the feed, yet such particles are enriched within the concentrate.


\subsection{Conditional univariate Tromp functions, conditioned on particle size and shape classes}\label{Sec:Univariate Tromp functions conditioned on particle size and shape classes}

To gain a deeper understanding of how particle morphology affects the separation behavior of particles, conditional univariate Tromp functions are computed, in addition to unconditioned bivariate Tromp functions considered in Equation~(\ref{Eq:Number weighted Tromp function}). This allows to analyze the influence of specific shape factors on the separation behavior across different size classes, e.g., how the aspect ratio of particles influences the separation behavior of small particles compared to that of larger ones. Additionally, studying the separation behavior of particles with respect to their size conditioned on specific shape classes, such as highly elongated particles compared to less elongated ones, provides even further insights into the particle separation process.

The computation of conditional univariate Tromp functions involves the need of determining conditional probability densities~\cite{Kun2017}.
For that purpose, the planar cross-section of a particle taken at random from a particle system (either concentrate or tailings) is represented by a random vector $(\AreaDiameter, \AspectRatio)$ with values in $[0,\infty) \times [0,1]$, where $\AreaDiameter$ denotes its random area-equivalent diameter and $\AspectRatio$ its random aspect ratio, with probability densities $f_{\AreaDiameter}$ and $f_{\AspectRatio}$, respectively. The joint probability density of $(\AreaDiameter, \AspectRatio)$ is denoted by $f_{\AreaDiameter, \AspectRatio}:\R^2\to[0,\infty)$. In the following, it is described how the conditional univariate Tromp function, conditioned on particle size classes,  can be computed. Using a similar approach, it is also possible to compute univariate Tromp functions conditioned on particle shape classes.

The conditional probability density $f_{\AspectRatio|\AreaDiameter\in[a,b]}:[0,1]\to[0,\infty)$ of $\AspectRatio$ provided that $\AreaDiameter\in[a,b]$ for some $a,b\ge 0$ with $a<b$ and $\Prob(\AreaDiameter\in[a,b])>0$ is  given by
\begin{linenomath}
\begin{equation}\label{Eq:Conditional probability density}
    f_{\AspectRatio|\AreaDiameter\in[a,b]}(\aspectRatio) = \frac{\int_a^b f_{\AreaDiameter,\AspectRatio}(\areaDiameter,\aspectRatio)\,\mathrm{d}\areaDiameter}{\int_a^bf_{\AreaDiameter}(\areaDiameter)\,\mathrm{d}\areaDiameter}\qquad\mbox{for each  $\aspectRatio\in[0,1]$.}
\end{equation}
\end{linenomath}
 Furthermore, the
 conditional univariate Tromp function $T_{\AspectRatio|\AreaDiameter\in[a,b]}:[0,1]\to[0,1]$ with respect to the random particle descriptor $\AspectRatio$, provided that $\AreaDiameter\in[a,b]$, is given by
\begin{linenomath}
\begin{equation}\label{Eq:Conditional Tromp function}
    T_{\AspectRatio|\AreaDiameter\in[a,b]}(\aspectRatio) = \begin{cases}\displaystyle
    \frac{\numberParticlesConcentrate_{|\AreaDiameter\in[a,b]}}{\numberParticlesFeed_{|\AreaDiameter\in[a,b]}}\,\frac{\densityconcentrate_{\AspectRatio|\AreaDiameter\in[a,b]}(\aspectRatio)}{\densityfeed_{\AspectRatio|\AreaDiameter\in[a,b]}(\aspectRatio)}, & \text{if } \densityfeed_{\AspectRatio|\AreaDiameter\in[a,b]}(\aspectRatio)>0,\\
    0, & \text{ if } \densityfeed_{\AspectRatio|\AreaDiameter\in[a,b]}(\aspectRatio)=0,
    \end{cases}
\end{equation}
\end{linenomath}
for each $\aspectRatio\in[0,1]$, where $\numberParticlesConcentrate_{|\AreaDiameter\in[a,b]}$ and $\numberParticlesFeed_{|\AreaDiameter\in[a,b]}$ denote the number of particles in the concentrate and feed with area-equivalent diameter in the interval of $[a,b]$, and $\densityfeed_{\AspectRatio|\AreaDiameter\in[a,b]}$ and $\densityconcentrate_{\AspectRatio|\AreaDiameter\in[a,b]}$ are the conditional probability densities of $\AspectRatio$ provided that $\AreaDiameter\in[a,b]$ for particles in the feed and concentrate, respectively. Here, both conditional densities  $\densityfeed_{\AspectRatio|\AreaDiameter\in[a,b]}$ and $\densityconcentrate_{\AspectRatio|\AreaDiameter\in[a,b]}$  are computed according to the formula given in Equation~(\ref{Eq:Conditional probability density}).

Note that the first factor on the right-hand side of Equation~(\ref{Eq:Conditional Tromp function}), i.e., the ratio of the number of particles in the concentrate and feed, conditioned on size or shape classes, cannot be directly obtained just like the yield in Equation~(\ref{Eq:Yield}) for unconditional bivariate Tromp functions. There is a lack of information on the mass of particles in different size or shape classes, making it challenging to directly compute the yield for conditional univariate Tromp functions. To address this problem, recall that $\numberParticlesConcentrate$ denotes the number of particles in the concentrate and that the probability that the cross-section of a random particle in the concentrate has an area-equivalent diameter $\AreaDiameter$ in the interval $[a,b]$ can be approximated by $\int_a^{b}\densityconcentrate_{\AreaDiameter}(\areaDiameter)\,\mathrm{d}\areaDiameter$, where 
$\densityconcentrate_{\AreaDiameter}$ denotes the corresponding probability density of particles in the concentrate. Then, the number of particles 
$\numberParticlesConcentrate_{|\AreaDiameter\in[a,b]}$
    in the concentrate can be approximated, whose cross-sections have an area-equivalent diameter in the interval $[a,b]$, by $\numberParticlesConcentrate_{|\AreaDiameter\in[a,b]}\approx \numberParticlesConcentrate\int_a^{b}\densityconcentrate_{\AreaDiameter}(\areaDiameter)\,\mathrm{d}\areaDiameter$. Analogously for the feed, one obtains $\numberParticlesFeed_{|\AreaDiameter\in[a,b]}\approx \numberParticlesFeed\int_a^{b}\densityfeed_{\AreaDiameter}(\areaDiameter)\,\mathrm{d}\areaDiameter$. Using these approximations, it is obtained that

\begin{linenomath}
\begin{equation}\label{Eq:Number ratio conditioned}
    \frac{\numberParticlesConcentrate_{|\AreaDiameter\in[a,b]}}{\numberParticlesFeed_{|\AreaDiameter\in[a,b]}} \approx  
    \frac{\numberParticlesConcentrate}{\numberParticlesFeed}\frac{\int_a^{b}\densityconcentrate_{\AreaDiameter}(\areaDiameter)\,\mathrm{d}\areaDiameter}{\int_a^{b}\densityfeed_{\AreaDiameter}(\areaDiameter)\,\mathrm{d}\areaDiameter},
\end{equation}
\end{linenomath}
where the yield $\numberParticlesConcentrate/\numberParticlesFeed$ is determined by means of Equation~(\ref{Eq:Yield}) and the values of the integrals on the right-hand side of~(\ref{Eq:Number ratio conditioned}) can be computed by numerical integration of  the marginal probability densities $\densityconcentrate_{\AreaDiameter}$ and $\densityfeed_{\AreaDiameter}$
obtained from fitting these probability densities to image measurements as described in Section~\ref{Sec:Univariate stochastic modeling of single particle descriptors}.

Taking into account the results, which have been obtained in~(\ref{Eq:Conditional probability density}) 
and~(\ref{Eq:Number ratio conditioned}), results in an approximation for the conditional Tromp function $T_{\AspectRatio|\AreaDiameter\in[a,b]}$
considered in Equation~(\ref{Eq:Conditional Tromp function}), i.e., 

\begin{linenomath}
\begin{equation}\label{Eq:Conditional Tromp function 2}
    T_{\AspectRatio|\AreaDiameter\in[a,b]}(\aspectRatio) \approx \begin{cases}\displaystyle
    \frac{\numberParticlesConcentrate}{\numberParticlesFeed}\,\frac{\int_a^b \densityconcentrate_{\AreaDiameter,\AspectRatio}(\areaDiameter,\aspectRatio)\,\mathrm{d}\areaDiameter}{\int_a^b \densityfeed_{\AreaDiameter,\AspectRatio}(\areaDiameter,\aspectRatio)\,\mathrm{d}\areaDiameter}, & \text{if } \int_a^b \densityfeed_{\AreaDiameter,\AspectRatio}(\areaDiameter,\aspectRatio)\,\mathrm{d}\areaDiameter>0,\\
    0, & \text{ if } \int_a^b \densityfeed_{\AreaDiameter,\AspectRatio}(\areaDiameter,\aspectRatio)\,\mathrm{d}\areaDiameter=0,
    \end{cases}
\end{equation}
\end{linenomath}
for each $\aspectRatio\in[0,1]$, where the 
bivariate probability densities $\densityfeed_{\AreaDiameter,\AspectRatio}$ and $\densityconcentrate_{\AreaDiameter,\AspectRatio}$ of particle descriptor vectors associated with particles in the feed and concentrate are computed as described in Section~\ref{Sec:Probability densities of descriptor vectors associated with particles in the feed and concentrate in flotation separation}.


\section{Results and discussion}\label{Sec:Results and discussion}

The results, which have been obtained by means of the methods described, are presented in the following sections. First, in Section~\ref{Sec:Classic flotation results}, typical flotation results in terms of grade and recovery as well as the mass and water pull are presented for the flotation experiments performed with  the \textit{MultiDimFlot} separation apparatus. Then, Sections~\ref{Sec:Influence of particle size and shape on the entrainment of ultrafine particles} and \ref{Sec:Influence of particle shape, size and wettability on the separation behavior of ultrafine particles} deal with bivariate Tromp functions to investigate the combined effects of particle size and shape on the separation. Finally, in Section~\ref{Sec:General discussion} the use of the MLA is discussed for determining the particle  descriptors in the case of the ultrafine particle fractions considered in this study.
Generally, it is assumed that there is no true flotation for hydrophilic particles, which in this case would apply to the completely hydrophilic systems ($\CZero$) consisting either of pristine glass spheres or fragments mixed with magnetite, where recovery is expected to occur via entrainment only. Flotation of the hydrophilic particle systems, therefore helps to study the influence of particle size and shape on their entrainment. Magnetite has a higher mass density, which results in a faster settling velocity compared to glass particles. Therefore, for the hydrophilic systems, it is expected that the recovery of glass particles by entrainment is higher, since the magnetite is drained back into the pulp more easily. The interaction of the gas bubbles and the glass fractions with increased hydrophobicity ($\CSix$ and $\CTen$) should result in more stable particle-bubble aggregates, hence they are recovered predominantly by true flotation, though, due to the fine particle sizes a certain degree of entrainment is expected for all particles.

\begin{figure}[ht!]
\captionsetup[subfigure]{labelformat=empty}
\begin{adjustwidth}{-\extralength}{-1.5cm}
        \centering
        \vspace{-0.5cm}
         \begin{subfigure}[b]{0.45\textwidth}
         \subcaption{~}
      \includegraphics[scale=0.4]{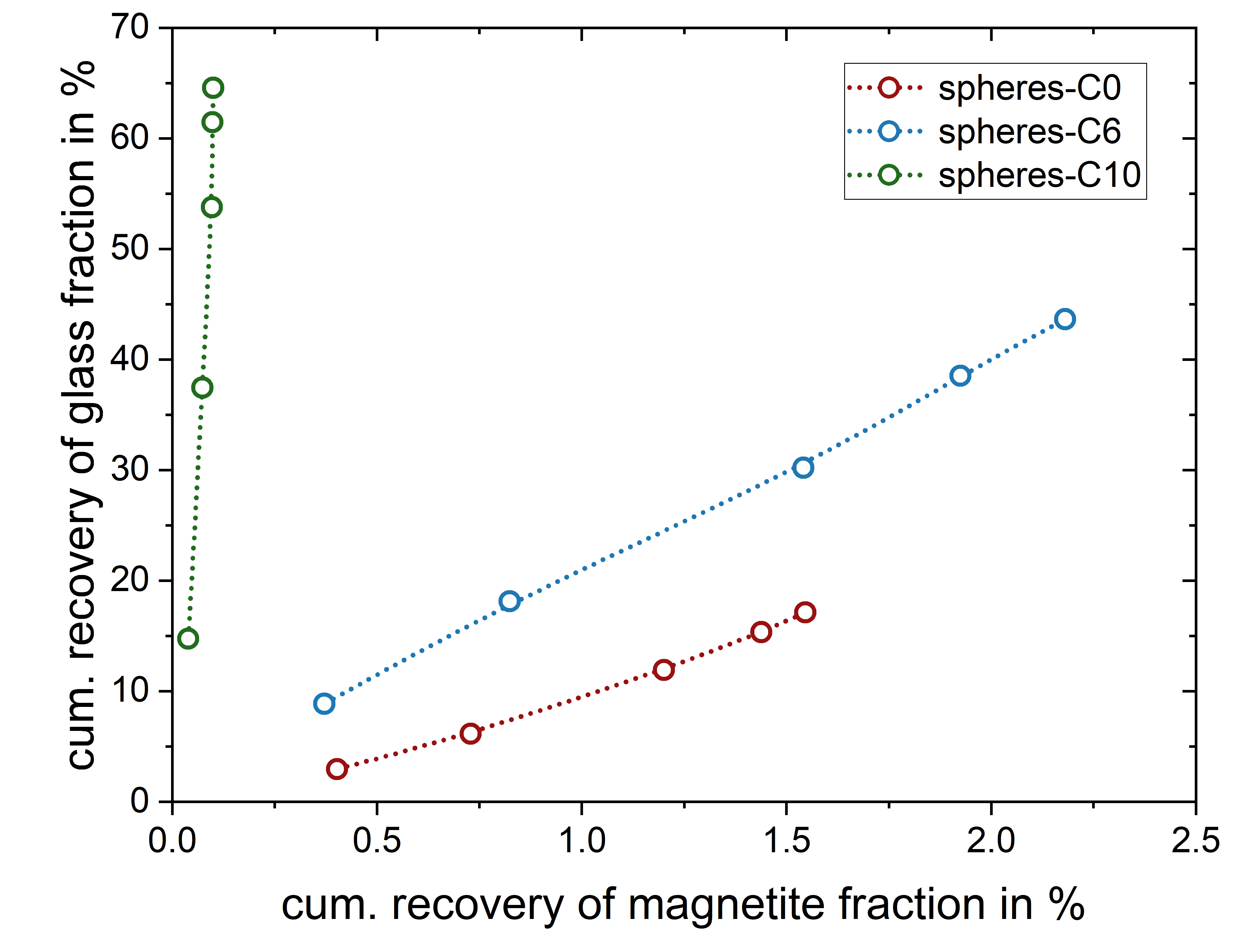}
     \end{subfigure}
     \begin{subfigure}[b]{0.45\textwidth}
     \subcaption{~}
     \includegraphics[scale=0.4]{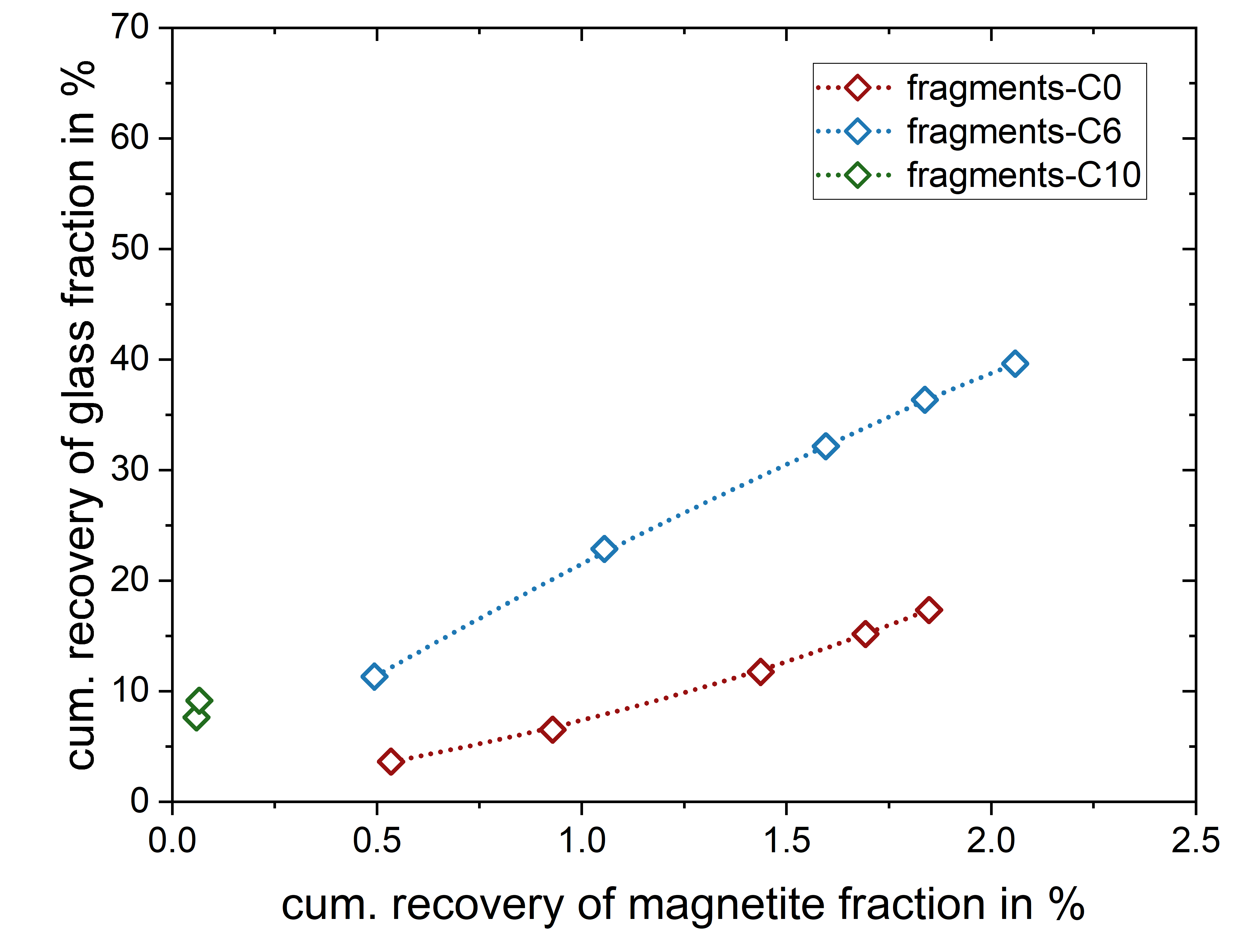}
     \end{subfigure}
     \end{adjustwidth}
    \caption{Fuerstenau upgrading curves showing the cumulative recovery of ultrafine glass particles, against the cumulative recovery of ultrafine magnetite, for the flotation experiments with \textit{MultiDimFlot}  using glass spheres (left) and fragments (right) with the three different wettability states: hydrophilic, $\CZero$ (red), moderately hydrophobic, $\CSix$ (blue) and strongly hydrophobic, $\CTen$ (green). The recovery is defined as the amount of recovered material in the concentrate in relation to its amount in the feed and is presented cumulatively over all concentrates.}
    \label{Fig:Fuerstenau}
\end{figure}


\subsection{Classic flotation results}\label{Sec:Classic flotation results}

The results of the flotation experiments are presented in Figure~\ref{Fig:Fuerstenau} as Fuerstenau-plots showing the cumulative recovery of the glass fractions versus the cumulative recovery of the magnetite fraction for the individual experiments using glass spheres and fragments with different wettability states.

\begin{figure}[ht!]
\captionsetup[subfigure]{labelformat=empty}
\begin{adjustwidth}{-\extralength}{-1.5cm}
        \centering
        \vspace{-0.5cm}
         \begin{subfigure}[b]{0.45\textwidth}
         \subcaption{~}
      \includegraphics[scale=0.4]{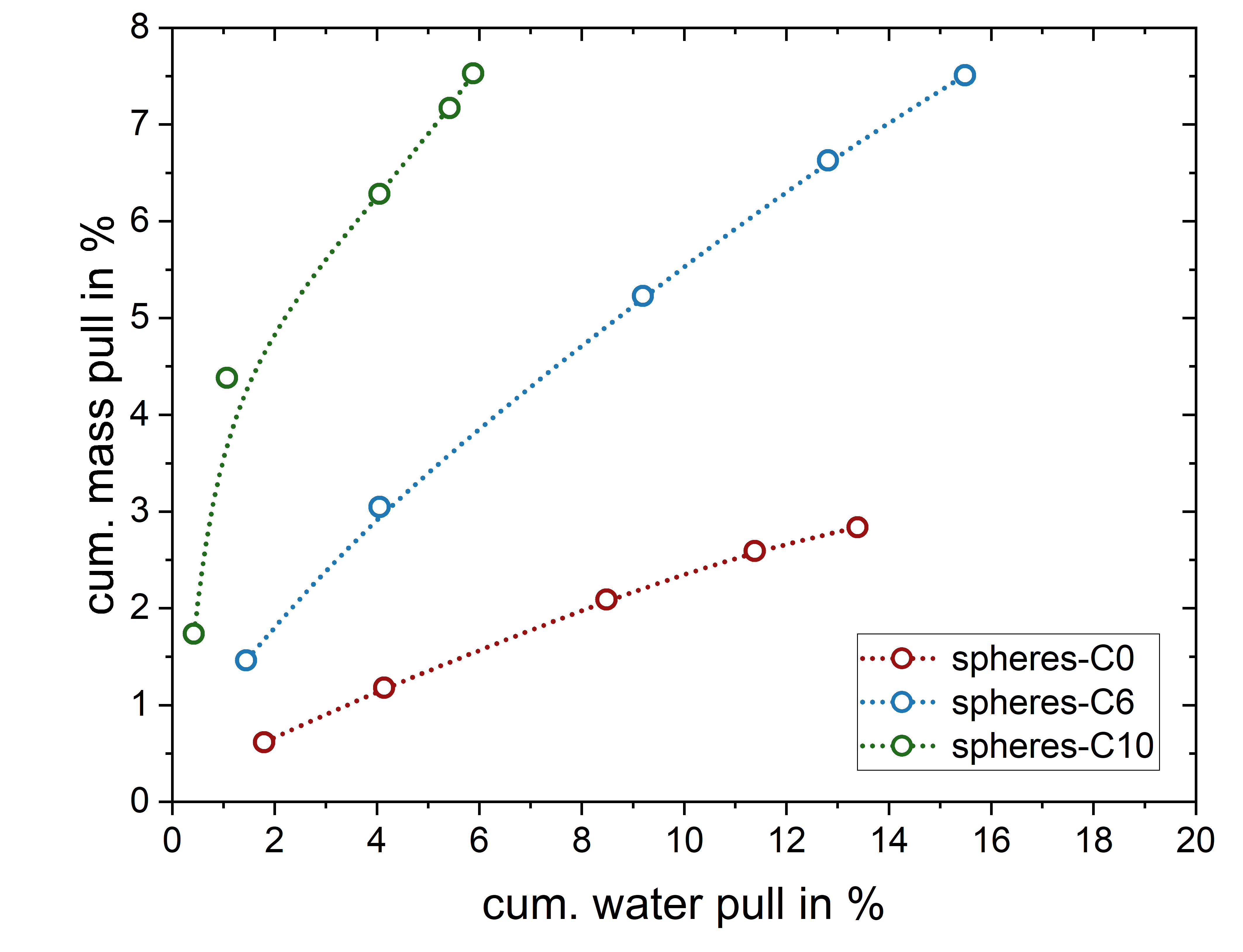}
     \end{subfigure}
     \begin{subfigure}[b]{0.45\textwidth}
     \subcaption{~}
     \includegraphics[scale=0.4]{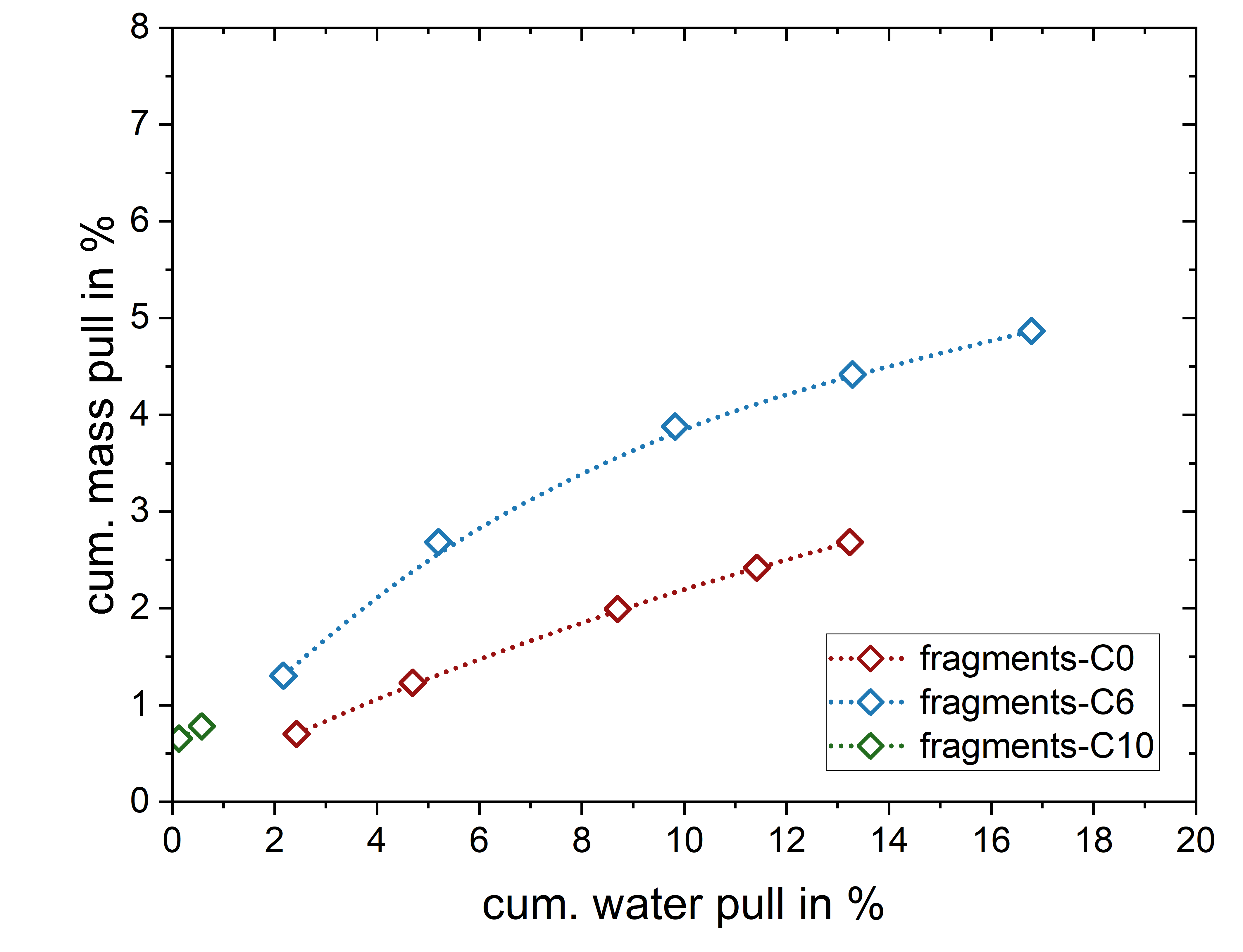}
     \end{subfigure}
     \end{adjustwidth}
    \caption{Cumulative mass versus water pull diagrams for the flotation experiments with \textit{MultiDimFlot} using glass spheres (left) and fragments (right) as the floatable fraction with the three different wettability states: hydrophilic, $\CZero$ (red), moderately hydrophobic, $\CSix$ (blue) and strongly hydrophobic, $\CTen$ (green). Ultrafine magnetite is used as the non-floatable fraction for all flotation experiments. The results are presented cumulatively over all concentrates.}
    \label{Fig:mass_water_pull}
\end{figure}

For all experiments it is apparent that the recovery of glass particles increases along with an increase in their hydrophobicity (from $\CZero$ via $\CSix$ to $\CTen$) and the glass particle recovery is much higher than that of magnetite, even for the hydrophilic ($\CZero$) fractions, where the effect of the particle mass density on the entrainment is visible. However, an influence of the particle shape on the entrainment cannot be observed, since the same amount of spheres and fragments (both $17 \%$) is entrained and also the total cumulative mass and cumulative water pull, which is displayed in Figure~\ref{Fig:mass_water_pull}, results in similar values for both particle systems (mass pull-spheres = $2.8 \%$, mass pull-fragments = $2.7 \%$ , water pull-spheres = $13 \%$, water pull-fragments = $13 \%$). This is an unexpected finding, since other studies showed that shape does have an effect on the particle entrainment, although the basis of comparison is not too high and the results of said studies are diverse. While Little et al.~\cite{Little2016} and Kupka et al.~\cite{Kupka2020} show that the entrainment is higher for round particles, Wiese et al.~\cite{Wiese2015} report the opposite for their particle system.

In comparison to the completely hydrophilic particle systems, the recovery of the moderately hydrophobic glass fractions ($\CSix$) is much higher, because they are supposed to be recovered mainly by true flotation, while the recovery of magnetite is not significantly increased.
The recovery of spheres is slightly higher than that of fragments (R-spheres-$\CSix=44 \%$ vs. R-fragments-$\CSix = 40 \%$), which is opposite to what would have been expected according to previous studies that observe higher recoveries for edgy and rough particles. The supposed reason for this is that rough particles rupture the liquid film, separating the bubble and the particle, more easily than spherical particles with a smooth surface, and therefore show higher probabilities of attachment and consequently also higher recoveries~\cite{Koh2009,Vaziri Hassas2016, Verrelli2014,Xia2017,Szczerkowska2018}. 
The opposite behavior is most probably due to two reasons which have been discussed by Sygusch et al.~\cite{Sygusch2023}: First, the set-up of the \textit{MultiDimFlot} separation apparatus with a combination of a turbulent pulp with a deep froth zone plays an important role. Especially the froth zone has a strong impact on the separation, however, most of the investigations regarding particle shape effects were conducted in set-ups where there was little to no froth and the recovery is dominated by effects in the pulp zone. Second, the investigated particle size fractions used in most studies, are much coarser than the ultrafine particles used in the present study, and it has been shown before that particle size significantly affects the flotation kinetics and the impact of particle shape (shape is more dominating for coarser particles, while roughness is more dominating for finer particles)~\cite{Rahimi2012}. 
Interestingly, while the mass pull of spheres is much higher than that of fragments (see Figure~\ref{Fig:mass_water_pull}), the water pull is very similar ($15.5 \%$ for spheres-$\CSix$, $16.8 \%$ for fragments-$\CSix$), indicating a significant difference in the froth characteristics, which are influenced strongly by particle properties. This effect is observed to an even larger extend for the particle systems with strongly hydrophobic glass particles ($\CTen$).  Here, the recovery of the spheres is up to $65 \%$, whereas only around $9 \%$ of fragments are recovered. 
During the flotation of strongly hydrophobic fragments extremely large coalescing bubbles were observed, complicating the scraping of the froth and which finally resulted in a complete froth collapse after $\SI{2}{\minute}$ of flotation, so that only two concentrates could be obtained, hence the low recovery for this case (experiments were repeated but were similar or even worse in outcome). Therefore, the results of the flotation experiments using strongly hydrophobic fragments cannot be compared directly to that of the others,  but it does show the complexity of the flotation process, since minor changes in the particle properties can have significant effects on the separation, as the only difference between the strongly hydrophobic spheres and fragments is their shape. Dipenaar~\cite{Dippenaar1982} reported that non-spherical particles can rupture the liquid film at smaller contact angles compared to spherical particles. Furthermore, Johansson et al.~\cite{Johansson1992} and Schwarz et al.~\cite{Schwarz2005} suggested that there might be an optimum contact angle not only for certain size fractions, but also for certain shapes, as further discussed by Sygusch et al.~\cite{Sygusch2023}.


\begin{figure}[h!]
    \centering
     \includegraphics[width=0.9\textwidth]{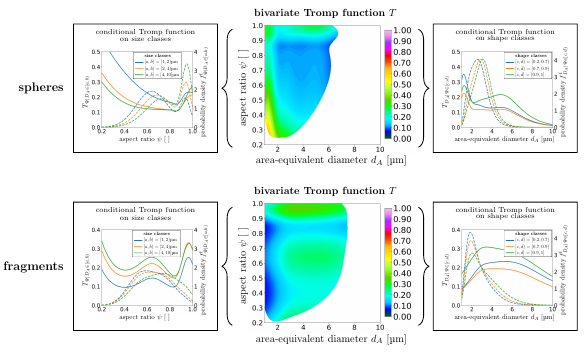}
    \caption{Bivariate Tromp functions for unesterifed hydrophilic spherical (upper row) and fragmented (lower row) glass particles (mixed with magnetite as feed for flotation experiments). Univariate Tromp functions conditioned on size and shape classes are visualized to the left and right of the corresponding bivariate Tromp function by solid lines. Univariate probability densities of area-equivalent diameter $\densityfeed_{\AreaDiameter|\AspectRatio\in[c,d]}$ and aspect ratio $\densityfeed_{\AspectRatio|\AreaDiameter\in[a,b]}$ conditioned on shape and size classes of glass particles in the feed, e.g., for small particles with area-equivalent diameter in the range of $[a,b) = [1,2)\,\SI{}{\micro\meter}$ or non-elongated particles with aspect ratio in the range of $[c,d]=[0.9,1]$, are illustrated by dashed lines, respectively.}
    \label{Fig:Bivariate and conditional Tromp function for glass particles}
\end{figure}

\subsection{Influence of particle size and shape on the entrainment of ultrafine particles}\label{Sec:Influence of particle size and shape on the entrainment of ultrafine particles}

Flotation experiments with purely hydrophilic particle systems allow for analyzing the influence of particle (size and shape) descriptor vectors  on the entrainment behavior of ultrafine particles. In the following, this kind of separation behavior of glass and magnetite particles is studied in order to gain a deeper understanding of the individual behavior of glass and magnetite, although, the flotation experiments are carried out using a mixture of glass (either spheres or fragments) and magnetite. First, bivariate Tromp functions regarding area-equivalent diameter and aspect ratio of planar cross-sections of differently shaped glass particles (spheres and fragments), are computed as described in Section~\ref{Sec:Stochastic modeling of particle descriptor vectors and computation of multivariate Tromp functions}, see Figure~\ref{Fig:Bivariate and conditional Tromp function for glass particles} (middle column). Colors ranging from yellow via red to pink indicate that particles in the feed with corresponding area-equivalent diameter and aspect ratio are more likely to be separated into the concentrate (i.e., with a probability larger than $0.5$), while colors ranging from dark green via blue to light green indicate that a particle with the corresponding descriptor vector is more likely to be separated into the tailings (i.e., particles with such particle descriptors have a probability smaller than $0.5$ to be separated into the concentrate). To provide a more meaningful interpretation of the probability that a particle is separated into the concentrate or the tailings, we computed the bivariate Tromp functions only for pairs $(\areaDiameter, \aspectRatio)$ of descriptor vectors corresponding to particles which are likely to be observed in the feed, where this set of pairs $(\areaDiameter, \aspectRatio)$ is indicated in white color in Figure~\ref{Fig:Bivariate and conditional Tromp function for glass particles}, see also~\cite{WilhelmSygusch2023}.

Additionally, conditional univariate Tromp functions of particle descriptors assigned to  glass particles and conditioned on size and shape classes are determined by means of Equation~(\ref{Eq:Conditional Tromp function}). These  Tromp functions are visualized in Figure~\ref{Fig:Bivariate and conditional Tromp function for glass particles} (left and right columns, solid lines). The corresponding conditional univariate probability densities $\densityfeed_{\AreaDiameter|\AspectRatio\in[c,d]}$ and $\densityfeed_{\AspectRatio|\AreaDiameter\in[a,b]}$ of area-equivalent diameter and aspect ratio assigned to planar cross-sections of either spheres or fragments are visualized by dashed lines for various intervals $[a,b]$ and $[c,d]$ which define size or shape classes on which the probability densities are conditioned.

For computing conditional univariate Tromp functions, particles are classified into three size classes: very fine particles with area-equivalent diameters between $1-\SI{2}{\micro\meter}$, medium-sized particles with area-equivalent diameters between $2-\SI{4}{\micro\meter}$, and large particles with area-equivalent diameters between $4-\SI{10}{\micro\meter}$. Similarly, particles are classified into three shape classes: highly-elongated particles with aspect ratios in the range between $0.2-0.7$, medium-elongated particles with aspect ratios between $0.7-0.9$, and almost non-elongated particles with aspect ratios larger than $0.9$.

Large variations are observed in the recovery probability of hydrophilic glass particles ($\CZero$) in Figure~\ref{Fig:Bivariate and conditional Tromp function for glass particles}, depending on their descriptor vectors. Furthermore, the bivariate Tromp functions display that differently shaped glass particle systems exhibit different separation behaviors (spheres vs. fragments).
For glass spheres, predominantly very fine particles with area-equivalent diameters of about $1-\SI{2}{\micro\meter}$ have a higher probability of reporting to the concentrate than coarser ones. This is observed over a wide range of aspect ratios for this size fraction, with slightly higher probabilities for smaller aspect ratios. On the other hand, for coarser particles, those with larger aspect ratio are more likely to be recovered. This is also observable in the univariate Tromp functions conditioned on size classes (Figure~\ref{Fig:Bivariate and conditional Tromp function for glass particles}, spheres, left-hand side), as the conditional univariate Tromp function for particles in the small particle class ($1-\SI{2}{\micro\meter}$) takes larger values for spheres with smaller aspect ratio, while for larger aspect ratios the probability is higher for coarser particle fractions. Similar results can be observed in the univariate Tromp functions conditioned on shape classes, as higher separation probabilities are reported for particles below $\SI{2}{\micro\meter}$ with low aspect ratio (0.2-0.7), whereas for coarser particles the conditional univariate Tromp function takes larger values for higher aspect ratios ($0.9-1$). Nevertheless, the conditional univariate Tromp functions show that the impact of particle size is more pronounced than that of particle shape for the case of ultrafine glass spheres. This behavior follows what has been reported in the literature cited above, i.e., entrainment increases with decreasing particle size, since the particles are more likely to be entrained in the froth lamella due to their low settling velocity.

In the case of glass fragments, a different outcome in the separation behavior is found. For this particle system, coarser fragments are more likely to be recovered than finer ones, which is contrary to what has been observed for glass spheres and to what is reported in the literature. This can be seen in both, the bivariate Tromp functions (Figure~\ref{Fig:Bivariate and conditional Tromp function for glass particles}, fragments, middle) and the univariate Tromp functions conditioned on shape classes (Figure~\ref{Fig:Bivariate and conditional Tromp function for glass particles}, fragments, right-hand side). Across all shape classes, the conditional univariate Tromp functions for fragments with a small area-equivalent diameter take smaller values compared to those for similarly sized spheres. Furthermore, while the recovery probability for the spheres generally decreases with increasing size, for the fragments it increases along with the particle size for all shape classes. Additionally, the values of univariate Tromp functions conditioned on size classes for fragments show variations for all size classes, depending on the considered aspect ratio, where the highest values are observed for very small and very large aspect ratios. This indicates that the influence of the particle shape on the separation behavior of the fragments is more significant than it is the case for the spheres.  
The observed differences in the recovery probabilities of spheres and fragments are particularly intriguing, since their cumulative recovery as well as the associated mass and water pull are quite comparable, see Section ~\ref{Sec:Classic flotation results}. Hence, an explanation for this behavior cannot be drawn from the classic non-particle-specific flotation evaluations.

\begin{figure}[h!]
     \includegraphics[width=0.9\textwidth]{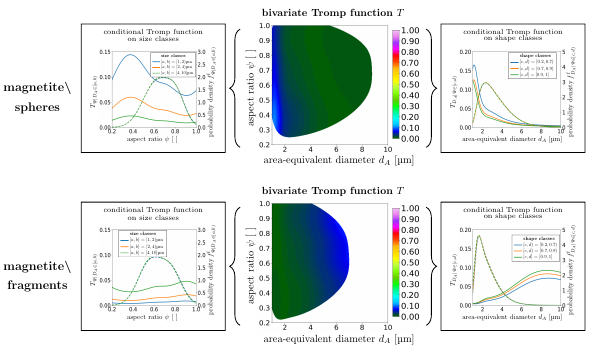}
    \caption{Bivariate Tromp functions for magnetite particles, mixed with unesterifed spherical (upper row) and fragmented (lower row) glass particles as feed. The most apparent dark green color corresponds to a probability of $T=0$. Univariate Tromp functions conditioned on size and shape classes are visualized to the left and right of the corresponding bivariate Tromp function by solid lines. The univariate probability densities  $\densityfeed_{\AreaDiameter|\AspectRatio\in[c,d]}$ and $\densityfeed_{\AspectRatio|\AreaDiameter\in[a,b]}$ of area-equivalent diameter and aspect ratio, conditioned on shape and size classes of magnetite particles in the feed, e.g., for small particles with area-equivalent diameter in the range of $[a,b) = [1,2)\,\SI{}{\micro\meter}$ or non-elongated particles with aspect ratio in the range of $[c,d]=[0.9,1]$, are visualized by dashed lines.}
    \label{Fig:Bivariate and conditional Tromp function for magnetite particles}
\end{figure}

Similar to the analysis of the influence of particle shape and size on the entrainment behavior for the hydrophilic glass fraction, the entrainment behavior of magnetite, as the non-floatable fraction, is investigated. Figure~\ref{Fig:Bivariate and conditional Tromp function for magnetite particles} displays  univariate as well as bivariate Tromp functions regarding area-equivalent diameter and aspect ratio of magnetite particles individually, but separated in a mixture with spherical (top row) and fragmented glass particles (bottom row). 

It can be observed from Figure~\ref{Fig:Bivariate and conditional Tromp function for magnetite particles}  that the behavior of the magnetite particles significantly differs when mixed with spheres compared to fragments. 
If magnetite is mixed with glass spheres, almost only magnetite with an area-equivalent diameter of about $1-\SI{2}{\micro\meter}$ is entrained, whereas only coarser magnetite particles with an area-equivalent diameter of about $3-\SI{7}{\micro\meter}$ are recovered when mixed with glass fragments. This strong dependency on the particle size is also observed in the univariate Tromp functions conditioned on shape classes (Figure~\ref{Fig:Bivariate and conditional Tromp function for magnetite particles},  right-hand side), as these conditional Tromp functions significantly differ depending on which glass particle system magnetite is mixed with.

 On the other hand, the shape of magnetite does not significantly influence the separation behavior, since only minor variations in univariate Tromp functions for different values of aspect ratios conditioned on size classes for magnetite (Figure~\ref{Fig:Bivariate and conditional Tromp function for magnetite particles}, left-hand side) are visible, although the conditional univariate Tromp functions take larger values for fine particles compared to larger size classes.
Based on these results, one could conclude that the differently shaped ultrafine particle fractions (spheres vs. fragments) show a significantly different entrainment behavior. Furthermore, a significant correlation between the behavior of magnetite and the hydrophilic glass particles is observed, since the magnetite behavior is strongly influenced by the particle fraction it is mixed with. However, one has to question the meaningfulness of the results. Therefore, the results obtained from image analysis are cross-checked via laser diffraction, though only the particle size can be evaluated with this technique. To compare both measurement methods, the probability densities derived from image measurements are transformed into volume-weighted probability densities of particle descriptor vectors. A comprehensive description of this transformation can be found in~\cite{WilhelmSygusch2023}. Laser diffraction measurements only provide probability densities of descriptor vectors associated to particles in mixtures of glass and magnetite. Without further assumptions, it is not possible to distinguish between glass and magnetite particles based on laser diffraction alone. In contrast, MLA measurements can distinguish between particles composed of different materials. To compare both measurement methods, the information on glass and magnetite particles obtained from image analysis is combined to derive volume-equivalent probability densities of area-equivalent diameter for the mixed particle fractions.

\begin{figure}[h!]
\captionsetup[subfigure]{labelformat=empty}
\begin{adjustwidth}{-\extralength}{-1.5cm}
        \centering
        \vspace{-0.5cm}
         \begin{subfigure}[b]{0.45\textwidth}
         \subcaption{~}
     \includegraphics[scale=1.9]{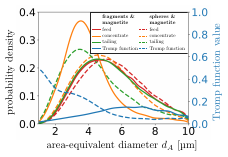}
     \end{subfigure}
     \begin{subfigure}[b]{0.45\textwidth}
     \subcaption{~}
           \includegraphics[scale=1.9]{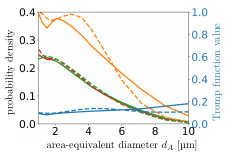}
     \end{subfigure}
     \end{adjustwidth}
    \caption{Univariate Tromp functions (blue) and volume-weighted probability densities of area-equivalent diameter for feed (red), concentrate (orange) and tailings (green) determined by means of image analysis (left) and laser diffraction (right) for mixtures of differently shaped glass particles (spheres and fragments) with magnetite. Solid lines represent results obtained for mixtures of fragments with magnetite, while dashed lines represent results determined for mixtures of spheres with magnetite. }
    \label{Fig:Comparison-Helos-MLA}
\end{figure}

Figure~\ref{Fig:Comparison-Helos-MLA} shows the probability densities of area-equivalent diameters associated to mixtures of glass and magnetite, obtained from image analysis on the left and from laser diffraction on the right. The resulting univariate Tromp functions computed from volume-weighted probability densities from image analysis of the mixture of glass spheres and magnetite match those of the number-weighed ones shown in Figures~\ref{Fig:Bivariate and conditional Tromp function for glass particles} and~\ref{Fig:Bivariate and conditional Tromp function for magnetite particles}, since rather fine particles are recovered more preferably than coarser ones. In the separation experiment with glass fragments, lower recovery probabilities for fines can be observed, however the increasing probability for coarser fragments is not as prominent in the volume-weighted probability densities.

 On the other hand, the univariate Tromp functions determined by means of laser diffraction measurements reveal a similar behavior for both mixtures and no clear influence of the particle size on the separation behavior is observed. The comparison of the univariate Tromp functions obtained from laser diffraction with those obtained from image analysis shows that there are some discrepancies between the results from different measurement techniques. This could be due to various reasons. 
First of all, the techniques themselves work in different ways. Particle size information obtained via laser diffraction refers to the sphere equivalent scattering intensities. On the other hand, image analysis offers insights into particle sizes via subsequent computation of particle descriptors from said images, i.e., planar surface sections (2D information), which usually leads to underestimated particle sizes. This would indeed result in slightly different results for particle sizes, but at least the resulting trend should be rather similar, which is not the case here. Possible errors could also come from the MLA approach, which includes representative sampling, the sample preparation (embedding the particles into an epoxy resin), and the measurement itself (with special focus on the limitation in resolution) as well as the final computation of the particle property descriptors from these images from which the univariate Tromp functions are derived. 
Another aspect that has to be considered is the statistical representativness, i.e., determining meaningful areas in the property space of the univariate Tromp function for which a sufficient number of particles with such properties is available. An indication of meaningful areas in separation functions can be obtained by examining conditional probability densities, such as those displayed in Figures~\ref{Fig:Bivariate and conditional Tromp function for glass particles} and \ref{Fig:Bivariate and conditional Tromp function for magnetite particles}. A reliable interpretation of the separation behavior can only be made for regions in which conditional probability densities have significantly large values for specific size and shape classes. For example, in the case of fragments, the probability densities of the feed show that particles with an area-equivalent diameter of around $1.5-\SI{3}{\micro\meter}$ and an aspect ratio of around $0.4-0.9$ are observed with large frequency and are therefore sufficiently relevant, whereas particles with properties outside of this range are not so frequently found which introduces uncertainties for the interpretation. For glass spheres, only particles in the feed with an area-equivalent diameter of around $2-\SI{4}{\micro\meter}$ and an aspect ratio of around $0.5-0.8$ and $0.9-1$, see Figure~\ref{Fig:Bivariate and conditional Tromp function for glass particles} top row, are observed with a sufficient frequency. Therefore the values of Tromp functions can be considered meaningful in such areas in the property space.
In conclusion, it is not called for to come up with a physical explanation of the phenomena right away, especially due to the fact that both hydrophilic components, i.e., glass ($\CZero$) and magnetite follow order and both show the distinct discrepancy in behavior comparing the glass sphere and the glass fragment system and their Tromp functions regrading size. However, due to the promising possibilities of the multivariate approach we decide to report on those findings here even though we cannot offer a conclusive explanation, which should be pursued in future research.


\subsection{Influence of particle size, shape and wettability on the separation behavior of ultrafine particles}\label{Sec:Influence of particle shape, size and wettability on the separation behavior of ultrafine particles}

\begin{figure}[ht!]
    \centering
     \includegraphics[width=0.9\textwidth]{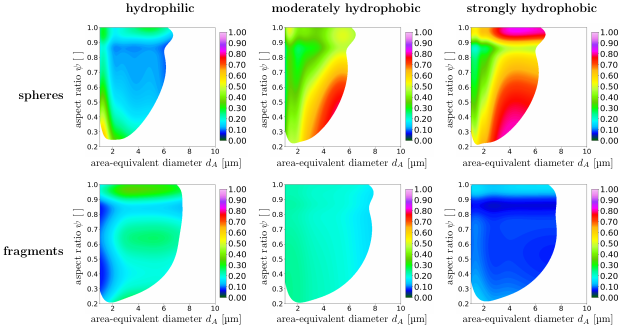}
    \caption{Bivariate Tromp functions for spherical (upper row) and fragmented (lower row) glass particles with different wettability states: unesterified hydrophilic (left column), moderately hydrophobic (middle column) and strongly hydrophobic (right column) glass particles.}
    \label{Fig:Tromp function for glass and different wettability modifications}
\end{figure}

In addition to flotation experiments with purely hydrophilic particle systems, further experiments were conducted with glass particles for which the wettability has been altered, i.e., the particles have been hydrophobized using two different n-alcohols to obtain different hydrophobicity states. In this way, not only the influence of shape and size can be investigated, but also the influence of wettability. The respective bivariate Tromp functions are visualized in Figure~\ref{Fig:Tromp function for glass and different wettability modifications}, showing the results for glass particles only (spheres in the upper row and fragments in the lower row), with hydrophobicity increasing from left to right, i.e., hydrophilic glass particles (left), moderately hydrophobic particles (middle), and strongly hydrophobic particles (right).

From the bivariate Tromp functions  in Figure~\ref{Fig:Tromp function for glass and different wettability modifications}, it can be seen that a change in wettability, i.e., an increase in hydrophobicity, influences the resulting recovery probabilities for both glass particle shapes, whereby the effect is observed to a larger extend for glass spheres (significantly larger values of bivariate Tromp functions). Here, the increase in hydrophobicity significantly increases the probability that the glass spheres are recovered in the concentrate, with the highest probabilities of up to $85 \%$ being observed for glass spheres with an aspect ratio of around $0.3-0.7$ and an area-equivalent diameter of around $3-\SI{6}{\micro\meter}$. If the hydrophobicity is increased further, the individual particle properties have a smaller impact on the separation behavior and the wettability is the dominating separation property. For both hydrophobic fractions of glass spheres recovery probabilities are higher for coarser particles. This would follow the expected behavior according to the literature cited above~\cite{Trahar1981,Schubert1996,Wills2016}, since hydrophobic particles are expected to be recovered via true flotation, i.e., they actually attach to the bubble and are recovered as a particle-bubble-aggregate in the froth, which is more efficient for coarser particles. However, a certain degree of entrainment is still expected for these ultrafine particle systems, which might account for the recovery probability of around $30 \%$ for the very fine glass spheres (below $\SI{2}{\micro\meter}$), which is similar to that of the purely hydrophilic particle systems.

\begin{figure}[ht!]
    \centering
     \includegraphics[width=0.9\textwidth]{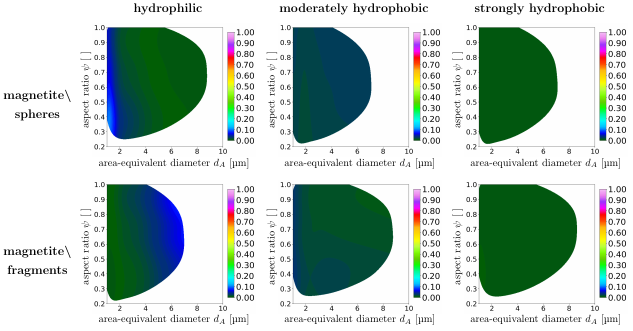}
    \caption{Bivariate Tromp functions for magnetite mixed with spherical (upper row) and fragmented (lower row) glass particles with different wettability states: unesterified hydrophilic (left column), moderately hydrophobic (middle column) and strongly hydrophobic (right column) glass particles. The most apparent dark green color corresponds to a separation probability of $T(x)=0$ for a descriptor vector $x\in[0,\infty)\times [0,1]$.}
    \label{Fig:Tromp function for magnetite and different wettability modifications}
\end{figure}

In the case of glass fragments, the  bivariate Tromp functions display a different outcome. While the entrainment probability of hydrophilic fragments is higher for coarser particles, the recovery probability for moderately hydrophobic fragments is slightly higher for finer particles than for coarser ones. Furthermore, no significant influence of the aspect ratio is observed for moderately hydrophobic fragments. Just as for their entrainment behavior (Section ~\ref{Sec:Influence of particle size and shape on the entrainment of ultrafine particles}), this is once more not in line with results from literature ~\cite{Trahar1981,Schubert1996,Wills2016}, since the recovery by true flotation should be more efficient for coarser particles, which in turn should result in higher probabilities in the bivariate Tromp functions for particles with a larger area-equivalent diameter. One reason for the higher recovery probability of fines could be, that the water pull, which correlates positively with the entrainment, of the moderately hydrophobic fragments is much higher than that of the moderately hydrophobic spheres ($17 \%$ vs. $8 \%$, respectively at a mass pull of $5 \%$). However, the water pull for hydrophilic fragments is even higher, but the bivariate Tromp functions show larger values for coarser particles, which is why this cannot be the sole reason. 
As already discussed in Section~\ref{Sec:Classic flotation results}, the results for the strongly hydrophobic glass fragments do not have the same basis of comparison, since very little mass pull was obtained due to the froth collapse. Here, the bivariate Tromp function shows very low separation probabilities for almost all particles, with slightly higher probabilities for fragments with a very high aspect ratio. 
Additionally to the results of glass particles, bivariate Tromp functions regarding area-equivalent diameter and aspect ratio are computed for the hydrophilic magnetite particles, see Figure~\ref{Fig:Tromp function for magnetite and different wettability modifications}, to investigate if their entrainment behavior is affected by the wettability variations of the glass particles they are mixed with. Here, it has to be kept in mind that only the wettability of the glass particles is changed, whereas the properties of magnetite are not modified in any way, i.e., any changes observed should be due to external factors. 

Figure~\ref{Fig:Tromp function for magnetite and different wettability modifications} shows that the entrainment probability of magnetite is generally rather low and its behavior is influenced by changes in the hydrophobicity of the glass particles. The most probable reason for this would be the different froth characteristics, which are strongly influenced by the properties of the particle system. This can be observed, for example, in the water pull, which is changing significantly with the feed used, as presented in Section~\ref{Sec:Classic flotation results}. While there is a certain influence of the particle size for the purely hydrophilic systems, when the magnetite is mixed with moderately hydrophobic glass spheres or fragments, this influence is not observed anymore since the entrainment probability is more or less the same across all ranges of the considered particle properties. If the magnetite is mixed with strongly hydrophobic glass particles, its entrainment probability is reduced to almost zero, which is most probably a result of the dry froth, i.e., a very low water pull, as less water is available that could drag along the ultrafine magnetite particles.


\subsection{Usability of MLA measurements to determine particle-discrete descriptor vectors}\label{Sec:General discussion}

\begin{figure}[ht!]
        \centering
    \includegraphics[width=0.75\textwidth]{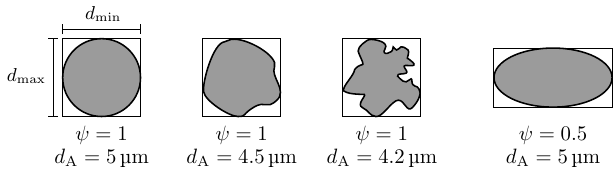}
    \caption{Illustration of size and shape descriptors. The area-equivalent diameter $\areaDiameter$ and aspect ratio $\aspectRatio$ of four different particles observed in planar sections are computed. Recall that the area-equivalent diameter $\areaDiameter$ of a particle's cross-section corresponds to the diameter of a circle with the same area, whereas the aspect ratio $\aspectRatio$ measures the elongation of the particle. In particular, small values of $\aspectRatio$  indicate elongated particles, whereas particles with aspect ratio close to $1$ are non-elongated.}
    \label{Fig:Illustration morphological descriptors}
\end{figure}

In this section, the usability of MLA measurements as a tool for evaluating the separation of particles by flotation, in particular in the case of ultrafines, is discussed.
First, it is important to note that the particle size and shape descriptors (area-equivalent diameter, aspect ratio) utilized in this study may not accurately represent the true 3D structure of the particles, as certain effects are not observable in 2D image data obtained by MLA. For instance, the aspect ratio of some particles can exhibit significant variations depending on the orientation of the particle with respect to the planar section imaged by MLA, introducing a potential bias for the distribution of aspect ratios determined from 2D data. Despite this limitation, these descriptors still provide a level of quantitative structural evaluation for assessing the impact of particle shape and size on their separation behavior. Figure~\ref{Fig:Illustration morphological descriptors} schematically visualizes the size and shape descriptors considered in this paper. 

The limitations of using the aspect ratio as a particle descriptor become apparent when examining  structural details of particles, as exemplified in Figure~\ref{Fig:Illustration morphological descriptors}  by the third particle from the left. This particular particle, despite having the same aspect ratio as the spherical particle, exhibits significant differences in surface roughness. Such structural nuances could play a crucial role in the attachment behavior of glass particles. Descriptors like the sphericity factor, roundness, or angularity~\cite{Furat2019, Zheng2015, Tafesse2013, Stachowiak2000} might offer a more suitable assessment of particle shape. However, these descriptors are not applicable in the present study due to the inability to accurately compute them from the available image data. Especially, for ultrafine particles involved in the present flotation separation experiments, obtaining the necessary high resolution for assessing shape descriptors poses a significant challenge using MLA measurements. Nevertheless, the elongation of particles should be considered when studying the influence of particle shape on separation behavior, as the particle systems of glass spheres and fragments significantly differ in particle aspect ratio and separation behavior, as discussed in Section~\ref{Sec:Influence of particle size and shape on the entrainment of ultrafine particles}.

\begin{figure}[ht!]
        \centering
      \includegraphics[width=0.75\textwidth]{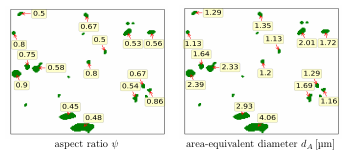}
    \caption{ False color cutouts of MLA measurements for glass fragments, which depict labeled regions that correspond to observed particles in the image data. For each particle its respective aspect ratio (left) and area-equivalent diameter (right) are highlighted. Particles with an area-equivalent diameter smaller than $\SI{1}{\micro\metre}$ are excluded in the analysis of MLA measurements. 
    }
    \label{Fig:Cutouts}
\end{figure}

Furthermore, the challenge of accurately characterizing the shape of particles, which could capture other structural nuances beyond elongation, is  highlighted in Figure~\ref{Fig:Cutouts}. Here, a cutout is shown featuring labeled regions of ultrafine glass fragments used in this study and imaged via MLA. Visual inspection already emphasizes the challenge of capturing fine nuances of particle shape beyond elongation. The corresponding size and shape descriptors of each particle are depicted. This offers a visual impression on different particle descriptor values extracted from image data, where it is important to note that particles with an area-equivalent diameter smaller than $\SI{1}{\micro\meter}$ are excluded from the analysis of MLA images.

Further limitations arise when working with images of ultrafine particles, particularly when extracting particles by the particle-based segmentation procedure described in Section~\ref{Sec:Particle-based segmentation} and consequently computing probability densities of particle descriptor vectors obtained from these segmentation results, as described in Section~\ref{Sec:Stochastic modeling of particle descriptor vectors and computation of multivariate Tromp functions}. In Figure~\ref{Fig:Bivariate probability densities of single fraction, feed and recomputed feed} bivariate probability densities of area-equivalent diameter and aspect ratio for glass spheres and fragments are presented that have been computed in three different ways. In particular, Figure~\ref{Fig:Bivariate probability densities of single fraction, feed and recomputed feed} (left column) shows the probability densities for glass particles obtained from analyzing individual fractions of said particles, i.e., glass particles prior to mixing with magnetite and the subsequent flotation experiments. On the other hand, Figure~\ref{Fig:Bivariate probability densities of single fraction, feed and recomputed feed} (middle column) displays the results obtained from analyzing the feed, i.e., by computing a bivariate probability density of descriptor vectors associated with glass particles observed in MLA data of glass-magnetite mixtures. Recall the numerical issues mentioned in Section~\ref{Sec:Probability densities of descriptor vectors associated with particles in the feed and concentrate in flotation separation} for computing bivariate Tromp functions with probability densities computed by these methods. Therefore, the probability densities corresponding to the reconstructed feed are computed as a convex combination of probability densities of descriptor vectors associated to glass particles in the concentrate and tailings by means of Equation~(\ref{Eq:Feed probability density as convex combination}), as detailed in Section~\ref{Sec:Probability densities of descriptor vectors associated with particles in the feed and concentrate in flotation separation}. These probability densities for glass spheres and fragments are visualized in the right column of Figure~\ref{Fig:Bivariate probability densities of single fraction, feed and recomputed feed}. For both types of glass particles (spheres, fragments), it can be observed that the resulting probability densities of particle descriptor vectors differ for all three procedures, with the largest variations being observed for the individual fractions.

\begin{figure}[ht!]
        \centering
    \includegraphics[width=0.9\textwidth]{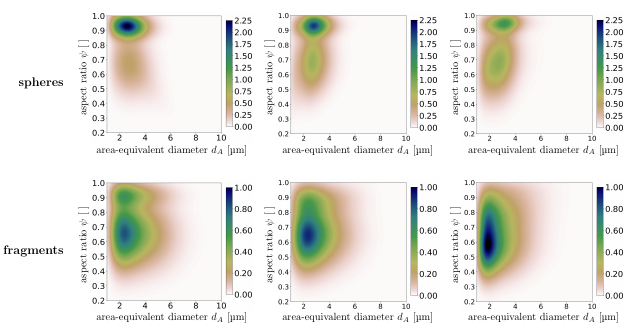}
    \caption{Bivariate probability densities of particle descriptor vectors consisting of area-equivalent diameter and aspect ratio associated with differently shaped glass particles (top: spheres and bottom: fragments). These densities are computed from MLA measurements of the individual fraction (left column) and the feed as mixed with magnetite (middle column), as well as recomputed from concentrate and tailings (right column), as detailed in Section~\ref{Sec:Probability densities of descriptor vectors associated with particles in the feed and concentrate in flotation separation}.}
    \label{Fig:Bivariate probability densities of single fraction, feed and recomputed feed}
\end{figure}

When analyzing the marginal probability densities of area-equivalent diameter and aspect ratio of the particle systems obtained by MLA data, these differences become even more evident, see  Figure~\ref{Fig:Marginal densities of single fraction, feed and recomputed feed}. This shows that the results obtained from the MLA images depend not only on the mathematical operations applied, but also on the particle system itself. Utilizing image measurements introduces a potential bias due to particles interacting with each other, possibly leading to agglomeration of particles of different materials and thus misclassification of particle shapes. This requires additional pre-processing steps for image data, to reduce the potential for errors in distinguishing between agglomerates and single particles, especially in ultrafine particle systems. Particularly, when applying a particle-based segmentation as described in Section~\ref{Sec:Particle-based segmentation} to a specific material without considering other materials, agglomeration between different materials is not taken into account. On the other hand, directly applying a particle-based segmentation to MLA measurements without first obtaining a phase-based segmentation results in a much larger amount of agglomerates observed in the image data. Segmentation of ultrafine particles within agglomerates poses a significant challenge, which is addressed  in Section~\ref{Sec:Particle-based segmentation}, but could be further improved to apply more advanced image pre-processing and segmentation tools. This indicates that the MLA measurements considered in this study may not accurately represent the particle systems, introducing a potential bias in the results regarding the analysis of separation behavior determined from image data in the case of ultrafine particles. Nevertheless, the results presented in this paper allow for a quantitative characterization of flotation results and could serve as a starting point for further investigations regarding the influence of particle morphology on the separation behavior.

\begin{figure}[ht!]
        \centering
    \includegraphics[width=\textwidth]{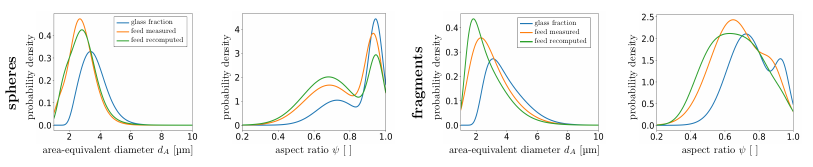}
    \caption{Marginal probability densities of area-equivalent diameter and aspect ratio associated with differently shaped glass particles (spheres, fragments). The densities are computed from MLA measurements of the individual fraction (glass fraction, blue), the feed as mixed with magnetite (feed measured, orange) and recomputed from concentrate and tailings (feed recomputed, green), as detailed in Section~\ref{Sec:Probability densities of descriptor vectors associated with particles in the feed and concentrate in flotation separation}.}
    \label{Fig:Marginal densities of single fraction, feed and recomputed feed}
\end{figure}



\section{Conclusion}\label{Sec:Conclusion}

This study addresses the application of bivariate Tromp functions as a tool for investigating the influence of  particle descriptor vectors of size, shape and wettability on the separation of ultrafine particles via froth flotation. Six different ultrafine feed systems were used in the separation experiments, in which the floatable fraction was either glass spheres or glass fragments with different levels of wettability, mixed with hydrophilic magnetite as the gangue material. Bivariate Tromp functions were computed based on MLA images of the feed and the flotation products and reveal quite some variations in the resulting recovery probabilities depending on the particle system used. 
The bivariate Tromp functions for glass spheres seem to be rather sensitive with respect to the investigated descriptors of particle size and shape as well as the induced change of hydrophobicity. On the other hand, the bivariate Tromp functions obtained for glass fragments show a rather different behavior. This is quite unexpected, since the results for grade and recovery as well as mass and water pull do not exhibit these kinds of variations. 
Investigations regarding the influence on the entrainment behavior of the hydrophilic glass systems show that the spheres follow the behavior commonly described in the literature, as their entrainment probability increases with decreasing particle size. Hydrophilic fragments, however, follow the opposite trend.
The use of laser diffraction to double-check the results obtained regarding the particle size influence did not confirm those obtained from MLA images. Reasons for this could be errors in the MLA measurement procedure, the 2D-limitation of MLA on imaging planar sections of particles, or the computation of the considered particle descriptor vectors from these images. Especially the latter is challenging in the case of ultrafine particle systems, as the MLA has limitations regarding the resolution. This also limits the accessibility of certain particle descriptors which might be more meaningful to investigate regarding their effect on flotation. 
Although the MLA might not be the most suitable method for analyzing ultrafines, the application of bivariate Tromp functions for the multidimensional evaluation of separation processes provides an innovative approach that helps to gain deeper insights into the particle behavior in such complex separation processes as froth flotation.


\vspace{6pt}

\authorcontributions{
Conceptualization, J.S., T.W., O.F., M.R. and V.S.; 
methodology, J.S., T.W.; 
software, T.W.; 
data curation, J.S. and K.B.; 
writing---original draft preparation, J.S., T.W. and O.F.; 
\mbox{writing---review} and editing, J.S., T.W., O.F., K.B., M.R. and V.S.; 
visualization, J.S., T.W.; 
supervision, M.R. and V.S.; 
project administration, M.R. and V.S.; 
funding acquisition, M.R. and V.S. All authors have read and agreed to the published version of the manuscript.
}

\funding{This research is partially funded by the German Research Foundation (DFG) via the research projects RU 2184/1-2, SCHM 997/27-2 and SCHM 997/45-1 within the priority programs SPP 2045 “Highly specific and multidimensional fractionation of fine particle systems with technical relevance” and SPP 2315 “Engineered artificial minerals (EnAM) -
A geo-metallurgical tool to recycle critical elements from waste streams”. 
}

\acknowledgments{The authors would like to thank Nora Stefenelli for conducting flotation experiments as part of her Bachelor thesis at the processing department of the Helmholtz Institute Freiberg for Resource Technology.}

\dataavailability{The datasets generated and/or analyzed during the current study are available from the corresponding authors on reasonable request.}

\conflictsofinterest{The authors declare no conflict of interest and the funders had no role in the design of the study; in the collection, analyses, or interpretation of data; in the writing of the manuscript; or in the decision to publish the~results.}

\begin{adjustwidth}{-\extralength}{0cm}

\reftitle{References}

\end{adjustwidth}

\begin{thebibliography}{999}


\bibitem[Trahar(1981)]{Trahar1981}
Trahar, W.J. A rational interpretation of the role of particle size in flotation. {\em International Journal of Mineral Processing} {\bf 1981}, {\em 8}, 289-327.

\bibitem[Schubert(1996)]{Schubert1996}
Schubert, H. \textit{Aufbereitung fester Stoffe, Band II: Sortierprozesse}, Deutscher Verlag für Grundstoffindustrie: Stuttgart, Germany, {\bf 1996}.

\bibitem[Wills(2016)]{Wills2016}
Wills, B.A.; Finch, J.A. \textit{Wills' Mineral Processing Technology: An Introduction to the Practical Aspects of Ore Treatment and Mineral Recovery}, Butterworth-Heinemann: Boston, USA, 2016.

\bibitem[Trahar(1976)]{Trahar1976}
Trahar, W.J.; Warren, L.J. The flotability of very fine particles—A review. {\em International Journal of Mineral Processing} {\bf 1976}, {\em 3}, 103–131.

\bibitem[Miettinen(2010)]{Miettinen2010}
Miettinen, T.; Ralston, J.; Fornasiero, D. The limits of fine particle flotation. {\em Minerals Engineering} {\bf 2010}, {\em 23}, 420–437.

\bibitem[DeGontijo(2007)]{DeGontijo2007}
De Gontijo, C.F.; Fornasiero, D.; Ralston, J. The limits of fine and coarse particle flotation. {\em Journal of Chemical Engineering} {\bf 2007}, {\em 85}, 739–747.

\bibitem[Konopacka(2010)]{Konopacka2010}
Konopacka, Z.; Drzymala, J. Types of particles recovery—Water recovery entrainment plots useful in flotation research. {\em Adsorption} {\bf 2010}, {\em 16}, 313–320.

\bibitem[Leistner(2017)]{Leistner2017}
Leistner, T.; Peuker, U.A.; Rudolph, M. How gangue particle size can affect the recovery of ultrafine and fine particles during froth flotation. {\em Minerals Engineering} {\bf 2017}, {\em 109}, 1–9.

\bibitem[Dai(2000)]{Dai2000}
Dai, Z.; Fornasiero, D.; Ralston, J. Particle-bubble collision models—A review. {\em Advances in Colloid and Interface Science} {\bf 2000}, {\em 85},  231–256.

\bibitem[Koh(2009)]{Koh2009}
Koh, P.T.L.; Hao, F.P.; Smith, L.K.; Chau, T.T.; Bruckard, W.J. The effect of particle shape and hydrophobicity in flotation. {\em International Journal of Mineral Processing} {\bf 2009}, {\em 93}, 128–134.

\bibitem[Vaziri Hassas(2016)]{Vaziri Hassas2016}
Vaziri Hassas, B.; Caliskan, H.; Guven, O.; Karakas, F.; Cinar, M.; Celik, M.S. Effect of roughness and shape factor on flotation characteristics of glass beads. {\em Colloids and Surfaces A: Physicochemical and Engineering Aspects} {\bf 2016}, {\em 492}, 88-99.

\bibitem[Verrelli(2014)]{Verrelli2014}
Verrelli, D.I.; Bruckard, W.J.; Koh, P.T.L.; Schwarz, M.P.; Follink, B. Particle shape effects in flotation. Part 1: Microscale experimental observations. {\em Minerals Engineering} {\bf 2014}, {\em 58}, 80-89.

\bibitem[Xia(2017)]{Xia2017}
Xia, W. Role of particle shape in the floatability of mineral particle: An overview of recent advances. {\em Powder Technology} {\bf 2017}, {\em 317}, 104-116.

\bibitem[Lu(2005)]{Lu2005}
Lu, S.; Pugh, R.J.; Forssberg, E. \textit{Interfacial Separation of Particles}. Elsevier: Amsterdam, The Netherlands, {\bf 2005}.

\bibitem[Chen(2022)]{Chen2022}
Chen, Y.; Zhuang, L.; Zhang, Z. Effect of particle shape on particle-bubble interaction behavior: A computational study using discrete element method. {\em Colloids and Surfaces A: Physicochemical and Engineering Aspects} {\bf 2022}, {\em 653}, 13003.

\bibitem[Kursun(2006)]{Kursun2006}
Kursun, H.; Ulusoy, U. Influence of shape characteristics of talc mineral on the column flotation behavior. {\em International Journal of Mineral Processing} {\bf 2006}, {\em 78}, 262–268.

\bibitem[Sygusch(2023)]{Sygusch2023}
Sygusch, J.; Stefenelli, N.; Rudolph, M. Ultrafine particle flotation in a concept flotation cell combining turbulent mixing zone and deep froth fractionation with a special focus on the property vector of particles. {\em Minerals}  {\bf 2023}, {\em 13}, 1099.

\bibitem[Little(2016)]{Little2016}
Little, L.; Wiese, J.; Becker, M.; Mainza, A.; Ross, V. Investigating the effects of particle shape on chromite entrainment at a platinum concentrator. {\em Minerals Engineering} {\bf 2016}, {\em 96–97}, 46–52.

\bibitem[Kupka(2020)]{Kupka2020}
Kupka, N.; Tolosana-Delgado, R.; Schach, E.; Bachmann, K.; Heinig, T.; Rudolph, M. R as an environment for data mining of process mineralogy data: A case study of an industrial rougher flotation bank. {\em Minerals Engineering} {\bf 2020}, {\em 146}, 106111.

\bibitem[Wiese(2015)]{Wiese2015}
Wiese, J.; Becker, M.; Yorath, G.; O’Connor, C. An investigation into the relationship between particle shape and entrainment. {\em Minerals Engineering} {\bf 2015}, {\em 83}, 211–216.

\bibitem[Albijanic(2010)]{Albijanic2010}
Albijanic, B.; Ozdemir, O.; Nguyen, A.V.; Bradshaw, D. A review of induction and attachment times of wetting thin films between air bubbles and particles and its relevance in the separation of particles by flotation. {\em Advances in Colloid and Interface Science} {\bf 2010}, {\em 159}, 1-21.

\bibitem[Drelich(2017)]{Drelich2017}
Drelich, J.W.; Marmur, A. Meaningful contact angles in flotation systems: Critical analysis and recommendations. {\em Surface Innovations} {\bf 2017}, {\em 6}, 19–30.

\bibitem[Johansson(1992)]{Johansson1992}
Johansson, G.; Pugh, R.J. The influence of particle size and hydrophobicity on the stability of mineralized froths. {\em International Journal of Mineral Processing}  {\bf 1992}, {\em 9}, 1–21.

\bibitem[Ata(2003)]{Ata2003}
Ata, S.; Ahmed, N.; Jameson, G.J. A study of bubble coalescence in flotation froths. {\em International Journal of Mineral Processing} {\bf 2003}, {\em 72}, 255–266.

\bibitem[Aveyard(1994)]{Aveyard1994}
Aveyard, R.; Binks, B.P.; Fletcher, P.D.I.; Peck, T.G.; Rutherford, C.E. Aspects of aqueous foam stability in the presence of hydrocarbon oils and solid particles. {\em Advances in Colloid and Interface Science} {\bf 1994}, {\em 48}, 93–120.

\bibitem[Dippenaar(1982)]{Dippenaar1982}
Dippenaar, A. The destabilization of froth by solids. I. The mechanism of film rupture. {\em International Journal of Mineral Processing}  {\bf 1982}, {\em 9}, 1-14.

\bibitem[Ata(2012)]{Ata2012}
Ata, S. Phenomena in the froth phase of flotation—A review. {\em International Journal of Mineral Processing}  {\bf 2012}, {\em 102-103}, 1–12.

\bibitem[Farrokhpay(2011)]{Farrokhpay2011}
Farrokhpay, S. The significance of froth stability in mineral flotation—A review. {\em Advances in Colloid and Interface Science}  {\bf 2011}, {\em 166}, 1–7.

\bibitem[Kaptay(2012)]{Kaptay2012}
Kaptay, G. On the optimum contact angle of stability of foams by particles. {\em Advances in Colloid and Interface Science}  {\bf 2012}, {\em 170}, 87–88.

\bibitem[Ata(2004)]{Ata2004}
Ata, S.; Ahmed, N.; Jameson, G.J. The effect of  hydrophobicity on the drainage of gangue minerals in flotation froths. {\em Minerals Engineering}  {\bf 2004}, {\em 17}, 897–901.

\bibitem[Schach(2019)]{Schach2019}
Schach, E.; Buchmann, M.; Tolosana-Delgado, R.; Lei{\ss}ner, T.; Kern, M.; {van den Boogaart}, K.G.; Rudolph, M.; Peuker, U.A. Multidimensional characterization of separation processes--Part 1: Introducing kernel methods and entropy in the context of mineral processing using SEM-based image analysis. {\em Minerals Engineering} {\bf 2019}, {\em 137}, 78-86.

\bibitem[Leissner(2016)]{Leissner2016}
Lei{\ss}ner, T.; Bachmann, K.; Gutzmer, J.; Peuker, U.A. MLA-based partition curves for magnetic separation. {\em Minerals Engineering} {\bf 2016}, {\em 94}, 94-103.

\bibitem[WilhelmSygusch(2023)]{WilhelmSygusch2023}
Wilhelm, T.; Sygusch, J.; Furat, O.; Bachmann, K.; Rudolph, M.; Schmidt, V. Parametric stochastic modeling of particle descriptor vectors for studying the influence of ultrafine particle wettability and morphology on flotation-based separation behavior. {\em Powders} {\bf 2023}, {\em 2}, 353-371.

\bibitem[Pereira(2021a)]{Pereira2021a}
Pereira, L.; Frenzel, M.; Hoang, D. H.; Tolosana-Delgado, R.; Rudolph, M.; Gutzmer, J. Computing single-particle flotation kinetics using automated mineralogy data and machine learning. {\em Minerals Engineering} {\bf 2021}, {\em 170}, 107054.

\bibitem[Pereira(2021b)]{Pereira2021b}
Pereira, L.; Frenzel, M.; Khodadadzadeh, M.; Tolosana-Delgado; R., Gutzmer; J. A self-adaptive particle-tracking method for minerals processing. {\em Journal of Cleaner Production} {\bf 2021}, {\em 279}, 123711.

\bibitem[Nelsen(2006)]{Nelsen2006}
Nelsen, R. \textit{An Introduction to Copulas}, Springer: Berlin/Heidelberg, Germany, {\bf 2006}.

\bibitem[Furat(2019)]{Furat2019}
Furat, O.; Lei{\ss}ner, T.; Bachmann, K.; Gutzmer, J.; Peuker, U.A.; Schmidt, V. Stochastic modeling of multidimensional particle characteristics using parametric copulas. {\em Microscopy and Microanalysis} {\bf 2019}, {\em 25}, 720-734.

\bibitem[Schubert(1989)]{Schubert1989}
Schubert, H. \textit{Aufbereitung Fester Mineralischer Rohstoffe, Band I}, Deutscher Verlag für Grundstoffindustrie: Stuttgart, Germany, 1989.

\bibitem[SyguschRudolph(2021)]{SyguschRudolph2021}
Sygusch, J.; Rudolph, M. A contribution to wettability and wetting characterisation of ultrafine particles with varying shape and degree of hydrophobization. {\em Applied Surface Science} {\bf 2021}, {\em 566}, 150725.

\bibitem[Heinig(2015)]{Heinig2015}
Heinig, T.; Bachmann, K.; Tolosana Delgado, R.; Van den Boogart, K.; Gutzmer, J. Monitoring gravitational and particle shape settling effects on MLA sample preparation, Proceedings of the 17th annual conference of the International Association of Mathematical Geosciences, Freiberg, Germany, {\bf 2015}, 200-206.

\bibitem[Fandrich(2007)]{Fandrich2007}
Fandrich, R.; Gu, Y:; Burrows, D.; Moeller, K. Modern SEM-based mineral liberation analysis. {\em International Journal of Mineral Processing} {\bf 2007}, {\em 84}, 310-320.

\bibitem[Schulz(2020)]{Schulz2020}
Schulz. B.; Sandmann, D.; Gilbricht, S. SEM-based automated mineralogy and its application in geo- and material sciences. {\em Minerals} {\bf 2020}, {10}, 1004.

\bibitem[Bachmann(2017)]{Bachmann2017}
Bachmann, K.; Frenzel, M.; Krause, J.: Gutzmer, J. Advanced identification and quantification of in-bearing minerals by scanning electron microscope-based image analysis. {\em Microscopy and Microanalysis} {\bf 2017}, {23}, 527-537.

\bibitem[Roerdink(2001)]{Roerdink2001}
Roerdink, J.B.T.M.; Meijster, A. The watershed transform: Definitions,
algorithms and parallelization strategies. {\em Fundamenta Informaticae} {\bf 2001}, {\em 41}, 187–228.

\bibitem[Soille(2003)]{Soille2003}
Soille, P. \textit{Morphological Image Analysis: Principles and Applications}, Springer: Berlin/Heidelberg, Germany, 2003.

\bibitem[Chiu(2003)]{Chiu2013}
Chiu, S.N.; Stoyan, D.; Kendall, W.S.; Mecke, J. \textit{Stochastic Geometry and its Applications}, J. Wiley \& Sons: Chichester, United Kingdom, 2013.

\bibitem[Furat(2018)]{Furat2018}
Furat, O.; Lei{ss}ner, T.; Ditscherlein, R.; Ond{\v{r}}ej, S.; Weber, M.; Bachmann, K.; Gutzmer, J.; Peuker, U.A.; Schmidt, V. Description of ore particles from X-Ray microtomography (XMT) images, supported by scanning electron microscope (SEM)-based
image analysis. {\em Microscopy and Microanalysis} {\bf 2018}, {\em 24}, 461--470.

\bibitem[Hilsenstein(2022)]{Hilsenstein2022}
\url{VolkerH/feret_diameter.py}. Available online: \url{https://gist.github.com/VolkerH/0d07d05d5cb189b56362e8ee41882abf} (accessed on 16 August 2022).

\bibitem[Held(2014)]{Held2014}
Held, L.; Bov{\'e}, D.S. \textit{Applied Statistical Inference: Likelihood and Bayes}, Springer: Berlin/Heidelberg, Germany, {\bf 2014}.

\bibitem[Akaike(1998)]{Akaike1998}
Akaike, H. Information theory and an extension of the maximum likelihood principle. {\em Selected Papers of Hirotugu Akaike} {\bf 1998}, 199-213.

\bibitem[Joe(2014)]{Joe2014}
Joe, H. \textit{Dependence Modeling with Copulas}, CRC Press: Boca Raton, FL, USA, {\bf 2014}.

\bibitem[Ditscherlein(2020)]{Ditscherlein2020}
Ditscherlein, R.; Furat, O.; {de Langlard}, M.; de Souza e Silva, J.M.; Sygusch, J.; Rudolph, M.; Lei{\ss}ner, T.; Schmidt, V.; Peuker, U.A. Multi-scale tomographic analysis for micron-sized particulate samples. {\em Microscopy and Microanalysis} {\bf 2020}, {\em 26}, 676-688.

\bibitem[Frank(2019)]{Frank2019}
Frank, U.; Wawra, S.E.; Pflug, L.; Peukert, W. Multidimensional particle size distributions and their application to nonspherical particle systems in two dimensions. {\em Particle \& Particle Systems Characterization} {\bf 2019}, {\em 36}, 1800554.

\bibitem[Buchmann(2018)]{Buchmann2018}
Buchmann, M.; Schach, E.; Tolosana-Delgado, R.; Leißner, T.; Astoveza, J.; Möckel, R.; Ebert, D.; Rudolph, M.; von de Boogaart, K.G.; Peuker, U.A. Evaluation of magnetic separation efficiency on a Cassiterite-Bearing Skarn ore by means of integrative SEM-based image and XRF–XRD data analysis. {\em Minerals} {\bf 2018}, {\em 8}, 390.

\bibitem[Kun(2017)]{Kun2017}
Kun II, P. \textit{Fundamentals of Probability and Stochastic Processes with Applications to Communications}, Springer: Berlin/Heidelberg, Germany, {\bf 2017}.

\bibitem[Szczerkowska(2018)]{Szczerkowska2018}
Szczerkowska, S.; Wiertel-Pochopien, A.; Zawala, J.; Larsen, E.; Kowalczuk, P.B. Kinetics of froth flotation of naturally hydrophobic solids with different shapes. {\em Minerals Engineering}  {\bf 2018}, {\em 121}, 90–99.

\bibitem[Rahimi(2012)]{Rahimi2012}
Rahimi, M.; Dehghani, F.; Rezai, B.; Aslani, M.R. Influence of the roughness and shape of quartz particles on their flotation kinetics. {\em International Journal of Mineral Processing}  {\bf 2012}, {\em 19}, 284–289.

\bibitem[Schwarz(2005)]{Schwarz2005}
Schwarz, S.; Grano, S. Effect of particle hydrophobicity on particle and water transport across a flotation froth. {\em Colloids and Surfaces A: Physicochemical and Engineering Aspects}  {\bf 2005}, {\em 256}, 157–164.

\bibitem[Xu(2001)]{Xu2001}
Xu, R. {\em Particle Characterization: Light Scattering Methods}, Springer: Berlin/Heidelberg, Germany, {\bf 2001}.

\bibitem[Merkus(2009)]{Merkus2009}
Merkus, H.G. \textit{Particle Size Measurements: Fundamentals, Practice, Quality}, Springer: Berlin/Heidelberg, Germany, {\bf 2009}.

\bibitem[Zheng(2015)]{Zheng2015}
Zheng, J.; Hryciw, R. Traditional soil particle sphericity, roundness and surface roughness by computational geometry. {\em Geotechnique} {\bf 2015}, {\em 65}, 494-506.

\bibitem[Tafesse(2013)]{Tafesse2013}
Tafesse, S.; Fernlund, J.M.R.; Sun, W.; Bergholm, F. Evaluation of image analysis methods used for quantification of particle angularity. {\em Sedimentology} {\bf 2013}, {\em 60}, 1100-1110.

\bibitem[Stachowiak(2000)]{Stachowiak2000}
Stachowiak, G.W. Particle angularity and its relationship to abrasive and erosive wear. {\em Wear} {\bf 2000}, {\em 241}, 214-219.


\end{thebibliography}
\end{document}